\documentclass[%
reprint,
superscriptaddress,
 amsmath,amssymb,
 aps,
 prb,
]{revtex4-2}

\usepackage{graphicx}
\usepackage{dcolumn}
\usepackage{bm}

\usepackage{graphicx}
\usepackage{dcolumn}
\usepackage{bm}
\usepackage[below]{placeins}
\usepackage{setspace} 
\usepackage{multirow}
\usepackage{float}
\usepackage{xcolor}

\newcommand{\FGT}{{Fe$_{3-x}$GeTe$_2$}}

\begin{document}

\title{Antiferromagnetic fluctuations and orbital-selective Mott transition in the van der Waals ferromagnet {{\FGT}}}

\author{Xiaojian Bai}
\email{baix@ornl.gov} \affiliation{Neutron Scattering Division,
Oak Ridge National Laboratory, Oak Ridge, TN 37831, USA}

\author{Frank Lechermann}
\email{frank.lechermann@ruhr-uni-bochum.de}
\affiliation{Institut f\"ur Theoretische Physik III, Ruhr-Universit\"at Bochum, D-44780 Bochum, Germany}

\author{Yaohua~Liu}
\affiliation{Neutron Scattering Division, Oak Ridge National
Laboratory, Oak Ridge, TN 37831, USA}

\author{Yongqiang Cheng}
\affiliation{Neutron Scattering Division, Oak Ridge National
Laboratory, Oak Ridge, TN 37831, USA}

\author{Alexander~I.~Kolesnikov}
\affiliation{Neutron Scattering Division, Oak Ridge National
Laboratory, Oak Ridge, TN 37831, USA}

\author{Feng Ye}
\affiliation{Neutron Scattering Division, Oak Ridge National
Laboratory, Oak Ridge, TN 37831, USA}

\author{Travis J. Williams}
\affiliation{Neutron Scattering Division, Oak Ridge National
Laboratory, Oak Ridge, TN 37831, USA}

\author{Songxue Chi}
\affiliation{Neutron Scattering Division, Oak Ridge National
Laboratory, Oak Ridge, TN 37831, USA}

\author{Tao Hong}
\affiliation{Neutron Scattering Division, Oak Ridge National
Laboratory, Oak Ridge, TN 37831, USA}

\author{Garrett E. Granroth}
\affiliation{Neutron Scattering Division, Oak Ridge National
Laboratory, Oak Ridge, TN 37831, USA}

\author{Andrew F. May}
\affiliation{Materials Science \& Technology Division, Oak Ridge National Laboratory, Oak Ridge, TN 37831, USA}

\author{Stuart Calder}
\email{caldersa@ornl.gov}
\affiliation{Neutron Scattering Division, Oak Ridge National
Laboratory, Oak Ridge, TN 37831, USA}

\date{\today}

\begin{abstract}
{\FGT} is a layered magnetic van der Waals material of interest for both fundamental and applied research. Despite the observation of intriguing physical properties, open questions exist even on the basic features related to magnetism: is it a simple ferromagnet or are there antiferromagnetic regimes and are the moments local or itinerant. Here, we demonstrate that antiferromagnetic spin fluctuations coexist with the ferromagnetism through comprehensive elastic and inelastic neutron scattering and thermodynamic measurements. Our realistic dynamical mean-field theory calculations reveal that the competing magnetic fluctuations are driven by an orbital selective Mott transition (OSMT), where only the plane-perpendicular $a_{1g}$ orbital of the Fe$(3d)$ manifold remains itinerant. Our results highlight the multi-orbital character in {\FGT} that supports a rare coexistence of local and itinerant physics within this material.

\end{abstract}

\maketitle

\begin{figure}[h]
\centering\includegraphics[width=1\columnwidth]{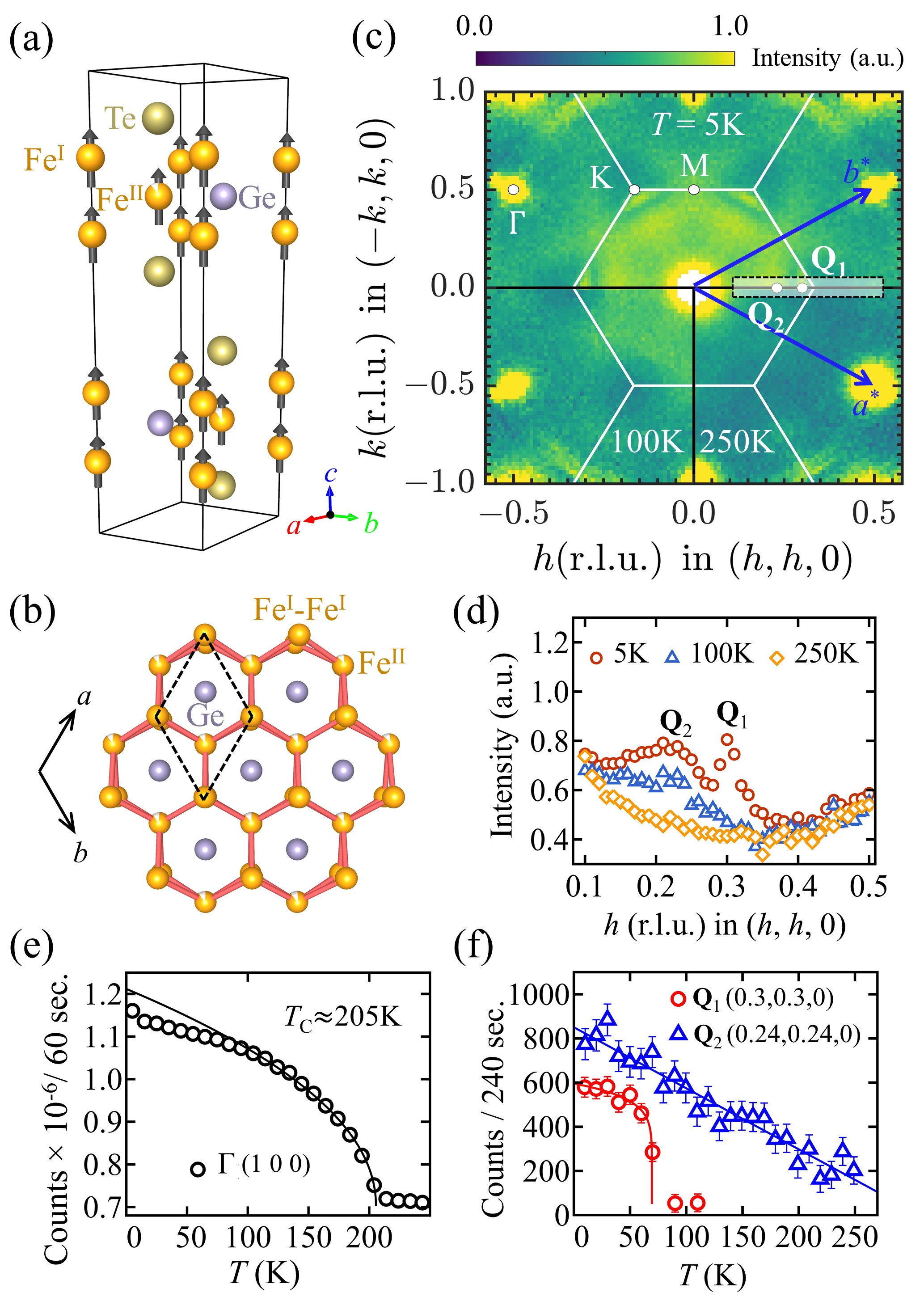}
\caption{\label{fig:diffraction} Elastic neutron scattering data showing FM and AFM static magnetic correlations in Fe$_{2.85}$GeTe$_2$. (a) Crystal structure and FM spin state of Fe$^\text{I}_2$Fe$^\text{II}_{1-x}$GeTe$_2$. (b) Single layer view, where the Fe atoms form a decorated honeycomb lattice. The dashed lines indicate the unit-cell. (c) Low-Q elastic scattering with $l$\,=\,$0.0\pm0.05$ r.l.u. The white lines are Brillouin zone boundaries. The shaded region indicates the line cuts along the $[110]$-direction integrated over $k$\,=\,$0.0\pm0.02$ r.l.u., shown in the panel (d). 
Intensity at the (e) FM zone center $\Gamma$ and (f) the AFM positions ${\bf Q}_1$ and ${\bf Q}_2$ as a function of temperature. Black and red curves are power-law fits of the data.  A linear background are fitted and removed for order-parameter data at ${\bf Q}_1$ [Fig.\,S5]. The paramagnetic background at 250\,K is removed at ${\bf Q}_2$ and the data offset by 200 for clarity.}
\end{figure}

Reducing the dimensionality of a compound to topologically constrained layers can enhance quantum phenomena and drive novel behavior. In this context two-dimensional (2D) layered materials that can exist from the bulk down to single layers due to weak interlayer van der Waals (vdW) bonding have undergone intense interest \cite{WOS:000448900900043,Adv_Mat}. The iron chalcogenide {\FGT} (FGT) (see Fig.\,\ref{fig:diffraction}(a)) has emerged as one of the central protagonists in 2D vdW material research. FGT is a rare example of a ferromagnetic (FM) metal vdW material, with the magnetism remaining robust down to the monolayer \citep{deng2018gate,liu2017wafer,fei2018two,roemer2020robust}, making it promising for device applications \cite{wang2019current,xu2019large}. Ionic gating has enhanced the magnetic ordering to room temperature in exfoliated flakes \citep{deng2018gate,zheng2020gate}. Additionally, there have been observations of anomalous Hall effect \citep{wang2017anisotropic,kim2018large,liu2018anomalous}, large anomolous Nernst effect with Berry curvature \cite{xu2019large}, and bubble and labyrinth domain structures and topologically protected skyrmions \citep{wu2020neel,ding2019observation}.

Despite intense studies, a deeper understanding of the physics in FGT, attributed to an apparent dichotomy of localized and itinerant electrons, has remained elusive. {This has led to debate on whether the magnetic ground state is a simple FM and how to form robust theoretical models.}
First of all, the Stoner exchange splitting is unlikely to be the sole driving force behind the magnetic ordering, instead, the interaction among localized moments may play an important role \citep{xu2020signature}.  Second, strong electronic correlations are essential to account for magnetic and thermodynamic properties \citep{zhu2016electronic}. Akin to iron-based superconductors \citep{tamai2010strong,mazin2008unconventional,georges2013strong}, the intra-atomic Hund coupling may be more relevant in this respect than close proximity to a Mott-critical regime \citep{corasaniti2020electronic}.
Furthermore, heavy-fermion behavior has been assigned to FGT based on experimental signatures below a characteristic temperature \citep{zhang2018emergence}, such as large mass renormalization \citep{chen2013magnetic} and Kondo screening \citep{zhao2021kondo}. 
Understanding magnetism in FGT and its conjunction with unique electronic properties renders an extensive investigation of energy- and momentum-resolved magnetic responses necessary.

In this Letter, we show that in FGT antiferromagnetic (AFM) spin fluctuations coexist with FM and explicate the behavior through an orbital selective Mott transtion (OSMT) that stabilizes both itinerant and local magnetism. This behavior is revealed experimentally through the static and dynamic magnetic response in neutron scattering measurements that shows continuous rod-like magnetic excitations emerging with characteristic wave-vectors centered around the K point.  We explain the underlying mechanisms with realistic dynamical mean-field theory (DMFT) calculations that reveal a rare OSMT occurs and drives the AFM behavior \citep{Good1982,AnisimovOSMT}. Our methodology leads to pinpointing the twofold-degenerate Fe-$e_g'$ orbitals of the Fe$(3d)$ shell as key behind both the magnetic transition and the OSMT. The multi-orbital character in FGT can explain the observations of local and itinerant physics and competing magnetism, as well as provide insights into potential Kondo behavior.

FGT crystallizes in the hexagonal space group P6$_3$/mmc \citep{deiseroth2006fe3gete2, chen2013magnetic, verchenko2015ferromagnetic, may2016magnetic}, containing two inequivalent Fe crystallographic sites, Fe$^\text{I}$ and Fe$^\text{II}$ in  Fig.\,\ref{fig:diffraction}(a) and (b). Depending on synthesis conditions, the vacancy concentration on the Fe$^\text{II}$ site can vary up to $30\%$  without changing the average crystal symmetry \citep{may2016magnetic}, while no apparent vacancy is found on the Fe$^\text{I}$ sites. A near-stoichiometric sample of FGT enters a FM phase with a strong $c$-axis anisotropy below the Curie temperature $T_\text{C}\approx$230\,K, which is suppressed with increasing vacancy concentration \citep{may2016magnetic}. 
The critical temperature can also be tuned by chemical doping \citep{drachuck2018effect,tian2019domain} and hydrostatic pressure \citep{wang2019pressure,ding2021ferromagnetism}.
Single crystals of {FGT} used in this study were synthesized using a self-flux method with a starting composition of Fe$_{6}$GeTe$_{9}$ and a maximum temperature of 1160\,$^{\circ}$C \citep{may2016magnetic, Drachuk2018}. A large single crystal of $\sim$1g with flat $c$-surfaces was selected [Fig.\,S3] and aligned in the $(hk0)$ scattering plane for all neutron experiments. A power-law fit to the temperature evolution of FM Bragg intensity yields a Curie temperature of $T_\text{C}$\,=\,$ 205(1)$\,K [Fig.\,\ref{fig:diffraction}(e)], consistent with a vacancy concentration of $x\sim 0.15$ \citep{may2016magnetic}. This Fe$_{2.85}$GeTe$_2$ composition therefore has significantly less vacancies than previous inelastic neutron studies \citep{calder2019magnetic}. See Supplementary Sec.\,S1 and S2 for further thermodynamic characterizations and experimental details. 

\begin{figure*}[tb]
\centering\includegraphics[width=0.85\textwidth]{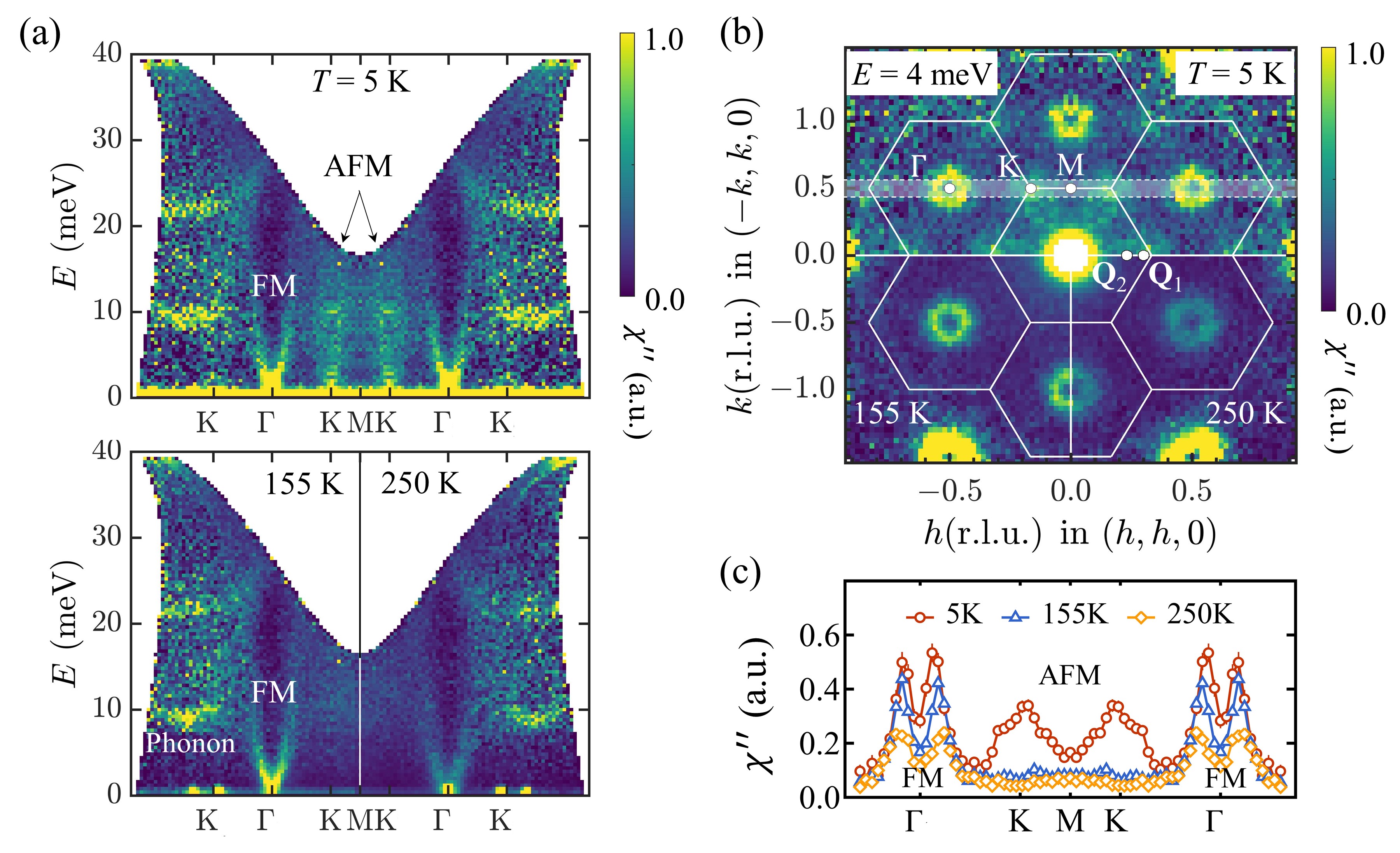}
\caption{\label{fig:inelastic} { Inelastic neutron-scattering data of Fe$_{2.85}$GeTe$_2$ showing coexistence of FM and AFM excitations. The Bose factor  is divided out in all panels, giving the imaginary part of the  dynamical susceptibility $\chi''$. 
(a) Temperature dependence of energy- and momentum-resolved excitations at $k$\,=\,$0.5\pm0.05$ r.l.u.~and $l$\,=\,$0.0\pm0.3$ r.l.u.. The optical modes around 10 and 20\,meV are attributed to phonons, see Sec.\,S6 for comparison with DFT-calculated
phonon spectra. (b) Low-Q constant-energy cut at $E$\,=\,$4\pm1$ meV. The white lines are Brillouin zone boundaries. (c) Line cuts along the [110]-direction at $k$\,=\,$0.5\pm0.1$ r.l.u. and $E$\,=\,$4\pm1$ meV.}}
\end{figure*} 

The elastic magnetic response is mapped out in Fig.\,\ref{fig:diffraction}(c) that uncovers AFM correlations developing at low temperatures within the previously observed FM phase. An orthogonal coordinate frame, \{$h{\bf a}^*+h{\bf b}^*$, $-k{\bf a}^*+k{\bf b}^*$, $l{\bf c}^*$\}, is used to present data in momentum space, with a generic momentum transfer labeled by ${\bf Q} = (h-k,h+k,l)$. At $T$\,=\,$5$\,K, two AFM features near the K point at ${\bf Q}_1 $\,=\,$ (0.3,0.3,0)$ and ${\bf Q}_2 $\,=\,$ (0.24,0.24,0)$ are observed [Fig.\,\ref{fig:diffraction}(c)-(d)]. Scattering at ${\bf Q}_1$ appears to be a weak Bragg reflection with an Ising-type order-parameter temperature dependence (critical exponent $\sim$\,$0.125$), which disappears above $T_\text{N}$\,=\,$70(1)$\,K [Fig.\,\ref{fig:diffraction}(f)]. ${\bf Q}_1$ has an anisotropic peak shape with a narrow width in the $[110]$-direction. The second feature is broad diffuse scattering near ${\bf Q}_2$ that persists to higher temperatures [Fig.\,\ref{fig:diffraction}(d)]. The intensity distribution remains mostly unchanged at $l$\,=\,$0$ and $0.5$ r.l.u.~[Fig.\,S4], reflecting a dominant $2$D character. Tracking the temperature dependence of scattering intensity at ${\bf Q}_2$ shows a linear decrease with increasing temperature up to room temperature.

\begin{figure*}[tb]
\centering\includegraphics[width=1\textwidth]{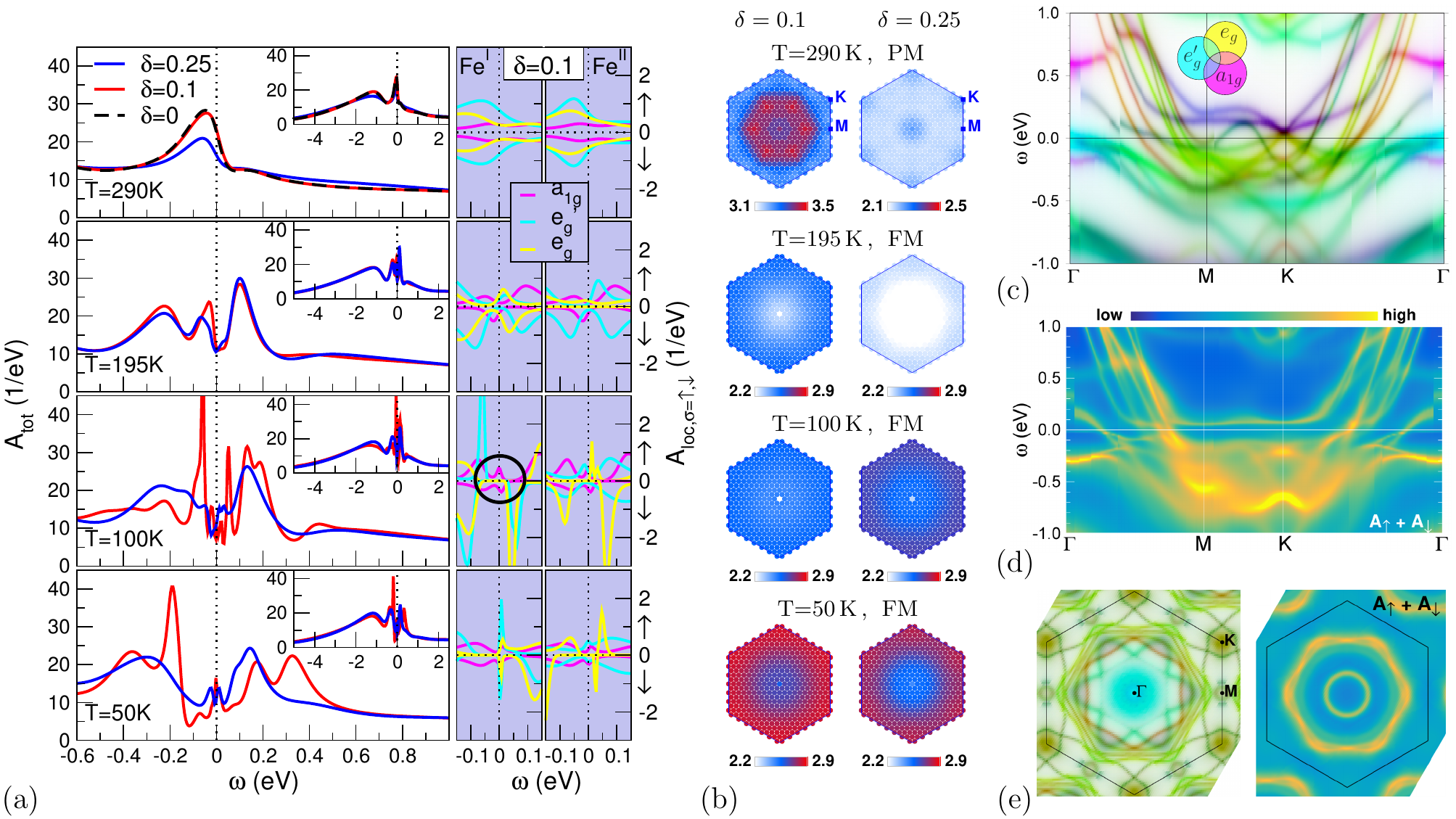}
\caption{DFT+DMFT results for Fe$_{3-\delta}$GeTe$_2$, where $\delta$\,=\,$x$ is the VCA doping level. (a) Total (left) and local Fe$(3d)$  (spin-)orbital-resolved (right) spectral function $A(\omega)$ 
for different $\delta$ and $T$. Left insets: larger energy window. Black circle for $T$\,=\,$100$\,K highlights the OSMT scenario. (b) $T$-dependent Lindhard spin susceptibility $\chi_s^{(0)}({\bf q},\omega$\,=\,$0)$ for $\delta$\,=\,$0.1,0.25$. (c,d) {\bf k}-resolved spectral-function properties for $\delta$\,=\,$0.1$. (c) $A({\bf k},\omega)$ along high-symmetry lines in a Fe$(3d)$ fatspec representation at $T$\,=\,$290$\,K. Mixed orbital weight appears as the accordingly mixed colors (see color scheme in graph). (d) Spin-summed $A({\bf k},\omega)$  for $T$\,=\,$50$\,K. (e) 
$k_z$\,=\,$0$ FS for $T$\,=\,$290$\,K in fatspec representation (left) and spin-summed intensity for $T$\,=\,$50$\,K (right).\label{fig:dmft} }
\end{figure*}

To investigate the dynamical signature of the observed magnetic correlations, we leverage inelastic neutron scattering on the large single crystal. The results are summarized in Fig.\,\ref{fig:inelastic}. As observed on measurements from samples with more Fe vacancies \citep{calder2019magnetic}, there are in-plane spin-waves consistent with FM ordering. These are well-resolved in momentum space below $\sim$8\,meV but significantly dampen as they enter the Stoner-like continuum at high energies. The magnetic signals may extend to above $100$\,meV energy transfer [Fig.\,S8]. No significant difference is observed for the low-energy FM spin-waves in the ordered phase between $5$ and $155$\,K after the Bose factor, $(1-\exp(-E/k_\text{B}T))^{-1}$, is removed [Fig.\,\ref{fig:inelastic}(c)]. As expected, they become more damped in the paramagnetic phase at $T$\,=\,$250$\,K. Systematic temperature evolution of the FM excitations are also observed in the out-of-plane directions [Fig.~S11]. Our linear spin-wave modeling suggests a rather small exchange coupling between Fe$^\text{I}$ and Fe$^\text{II}$ sites, differ from earlier neutron-scattering studies \citep{calder2019magnetic, trainer2021relating} and DFT calculations \citep{deng2018gate,jang2020origin}.

Of most significance, the inelastic neutron data reveals an AFM dynamical response in FGT, in accordance with the measured AFM elastic signals. Continuous rod-like excitations emerge near the K point of the hexagonal Brillouin zone at $T$\,=\,$5$\,K [Fig.\,\ref{fig:inelastic}(a)]. Strikingly, they are removed at $T$\,=\,$155$\,K well below the Curie temperature $T_\text{C}$\,=\,$205(1)$\,K, in sharp contrast with the FM spin-waves, indicating previously unseen competing AFM interactions. See Fig.\,S6 and S7 for more details on the temperature dependence. A close inspection of the low-Q region in a constant energy cut at $E$\,=\,$4\pm1$\,meV, [Fig.\,\ref{fig:inelastic}(b)], reveals that the inelastic signals extend from the K point to the $\Gamma$ point, covering both AFM ${\bf Q}_1$ and ${\bf Q}_2$ positions observed in the elastic channel. Data collected with a lower incident neutron energy $E_\text{i}$\,=\,$25$\,meV shows that the signal is gapless within the instrumental resolution of $0.65$\,meV [Fig.\,S8]. Applying a magnetic field of $\mu_0H$\,=\,$4$\,T along the $c$-axis did not yield clear changes in either FM and AFM magnetic excitations [Fig.\,S9], therefore it is likely that the dominant effect of the field is to align the ferromagnetic domains without changing microscopic magnetic correlations. 

To provide insight into the mechanisms driving the coexisting magnetism observed experimentally we performed charge self-consistent DFT+DMFT \citep{Savrasov01,Pourovskii07,Grieger2012} calculations, and the results are summarized in Fig.\,\ref{fig:dmft}. Finite doping is realized by the virtual-crystal approximation (VCA) originating from the Fe$^\text{II}$ site. The Fe$(3d)$ orbitals define the correlated subspace, with Hubbard $U$\,=\,$5$\,eV and Hund-rule coupling $J_{\rm H}$\,=\,$0.7\,$eV, in line with previous calculations \citep{Zhu2016}. Due to the hexagonal symmetry, the respective Fe five-fold states split into three classes; a $d_{z^2}$-like $a_{1g}$ orbital, as well as two degenerate $e_g'$ and two degenerate $e_g$ orbitals (for more details see Sec.\,S4).

Below the FM transition, $T$\,=\,$195$\,K in Fig.\,\ref{fig:dmft}(a), the electronic spectral weight at low energy is reduced and shifted to sidebands at $\sim\pm0.2$\,eV. Right at the Fermi level $\varepsilon_{\rm F}$ a pseudogap(-like) regime becomes visible. For $T$\,=\,$100$\,K this regime becomes more pronounced and an OSMT in the $\{e_g,e_g'\}$ orbital sector has occured, whilst the $a_{1g}$ sector remains metallic. Sharp low-energy $\{e_g,e_g'\}$ resonances are reappearing at even lower $T$, here shown at $T$\,=\,$50$\,K, suggesting a Kondo coupling between the $\{e_g,e_g'\}$ localized states and the itinerant $a_{1g}$ orbitals. The OSMT physics is most strongly realized on the Fe$^\text{I}$ sites and becomes weaker for the larger doping $\delta$\,=\,$0.25$. 

The OSMT-driven physics leads to a specific correlation-induced contribution to the local-moment formation in FGT. And this contribution gives rise to emerging AFM fluctations known for Mott-critical systems. The ${\bf q}$-dependent spin susceptibility $\chi_s^{(0)}$ in Fig.\,\ref{fig:dmft}(b) shows the growth of these AFM fluctuations (rising intensity at the zone boundary) with lowering $T$, as also observed experimentally (cf. Fig.\,\ref{fig:inelastic}). Further in agreement with the experimental data, the amplitude is somewhat larger around the K point than at the M point (see Sec.\,S4). The ${\bf k}$-resolved features of the correlated electronic structure at low energy are visualized in Fig.\,\ref{fig:dmft}(c) for ambient $T$ with a fatspec representation, i.e. spectral weight colored according to the respective Fe$(3d)$ orbital weight. 
Close to $\Gamma$ the $e_g'$ orbitals show flattened dispersion, which may be part of the root for the FM instability. A Dirac(-like) crossing point with substantial $a_{1g}$ weight is located at K because of the hexagonal in-plane lattice structure.  

The multi-sheet interacting Fermi surface (FS) displays a rather intricate topology. This complexity is reduced in the low-$T$ phase at 50\,K (see Fig.\,\ref{fig:dmft}(d)), where two dominant FS sheets around $\Gamma$ are established. Note that the outer sheet actually represent two entangled sub-sheets. Subtle states are encountered right at $\Gamma$ and at K, but a sharp single-level feature near $\Gamma$ may connect to Kondo physics. Flat dispersions further above and below the Fermi level can also be observed along M-K. The interacting FS and the additionally revealed ${\bf k}$-dependent features at low energy are in good agreement with angle-resolved photoemission data \citep{zhang2018emergence}.
The Fe ordered moments (see Sec.\,S4) become smaller by $\sim$20\% for $\delta$\,=\,$0.25$, in line with the experimental trend \citep{may2016magnetic}. This reduction with higher hole doping may be attributed to the parallel decrease of the OSMT strength.

Concerning the origin of the OSMT in FGT, differences in the respective orbital-resolved Fe$(3d)$ fillings and dispersions seem most crucial. The Fe$^\text{I}$-$a_{1g}$ orbital shows a pronounced bonding-antibonding splitting and is most-itinerant with electron filling $n_{a_{1g}}\sim 1.5$, whereas the $e_g'$ orbitals become integer-filled with $n_{e_g'}\sim 3$ in the interacting regime. The filling of the $e_g$ orbitals is nominally somewhat below three electrons, but their more ligand-hybridized character renders an obvious site distinction difficult. The OSMT is thus driven from the $e_g'$ sector and $e_g$ seemingly locks in. Hence interestingly, the $e_g'$ orbitals are apparently the key behind both the FM transition {\sl and} OSMT. On the more ligand-affected Fe$^\text{II}$ site, the strong orbital differentiation is smeared out, also due to the direct onsite vacancies.
Note that the present five-orbital OSMT scenario with three electrons in the twofold-degenerate $e_g'$ orbitals has to involve the Hund $J_{\rm H}$ in a more subtle manner than in conventional OSMT candidates, such as ruthenates \cite{AnisimovOSMT}. This may also be inferred from the small $\sim100-200$\,meV charge gap obtained here for the localized $\{e_g',e_g\}$ states.

Considering the experimental and theoretical results presented here on FGT allows new insights into its exotic and diverse behavior. The OSMT naturally explains the coexistence of localized and itinerant electrons in strongly correlated FGT by providing a multi-orbital character to separately host these behavior. Moreover, initially a general connection between OSMT and heavy-fermion physics was made in Ref.~\citep{demedici05}. While the connection may be subtle \citep{Pepin2007,deleo08,VojtaOSMT}, an example is found in the direct fitting for the $f$-electron material UPt$_3$ \citep{zwicknagel2002}. The orbital-selective scenario revealed here for FGT provides a natural origin for the measured heavy-fermion signatures \citep{zhang2018emergence,chen2013magnetic,zhao2021kondo}. Kinetic-exchange within this Mott-critical subspace then drives nearest-neighbor AFM fluctuations, which manifests in the spin-susceptibility enhancement at the BZ boundary observed experimentally and theoretically here and debated in the literature. Looking forward, as the low-temperature DFT+DMFT treatment of the realistic system is hindered by computational limitations, further details and cutting-edge data on the OSMT-based interplay between Kondo screening and magnetic order/fluctuations have to be addressed in tailored model-Hamiltonian studies.

In conclusion, an orbital-selective Mott transition has been shown to drive the emergent properties in {FGT}. This provides a singular multi-orbital character to this material, which both reconciles the apparent dual nature of local and itinerant magnetism and explains the observation of AFM fluctuations from the presented neutron-scattering data. Unexpected signatures of heavy fermion physics in previous studies of {FGT} have proven to be challenging to rationalize, however the uncovering of OSMT physics provides a clear route for the solution of this problem. The results presented here represent a significant advancement in understanding the coexistence of itinerant and local moments in a canonical quasi-2D vdW ferromagnetic material and may have relevant consequences for spin and orbital dependent electronic functions within wider spintronic and topological transport research.


\addtocontents{toc}{\protect\setcounter{tocdepth}{1}}
\begin{acknowledgments}
X.B. thanks Cristian Batista for valuable discussions. Sample synthesis and characterization (A.F.M.) was supported by the U.S. Department of Energy, Office of Science, Basic Energy Sciences, Materials Sciences and Engineering Division. This research used resources at the High Flux Isotope Reactor and Spallation Neutron Source, a DOE Office of Science User Facility  operated  by  the  Oak  Ridge  National  Laboratory. F.L. acknowledges support from the European XFEL and the Center for Computational Quantum Physics of the Flatiron Institute under the Simons Award ID 825141. Computations were performed at the JUWELS Cluster of the J\"ulich Supercomputing Centre (JSC) under project numbers hhh08 and miqs.
\end{acknowledgments}

\bibliographystyle{apsrev4-2}
\bibliography{references}

\addtocontents{toc}{\protect\setcounter{tocdepth}{0}}

\clearpage
\onecolumngrid
\setcounter{figure}{0}
\setcounter{table}{0}
\setcounter{equation}{0}
\setcounter{section}{0}
\renewcommand{\thefigure}{S\arabic{figure}}
\renewcommand{\thetable}{S\arabic{table}}
\renewcommand{\theequation}{S\arabic{equation}}
\renewcommand{\thesection}{S\arabic{section}}   

\begin{center}
{\large \bf Supplementary Information}
\end{center}

\section{Thermodynamic characterization}
\begin{figure*}[h]
\centering\includegraphics[width=0.9\textwidth]{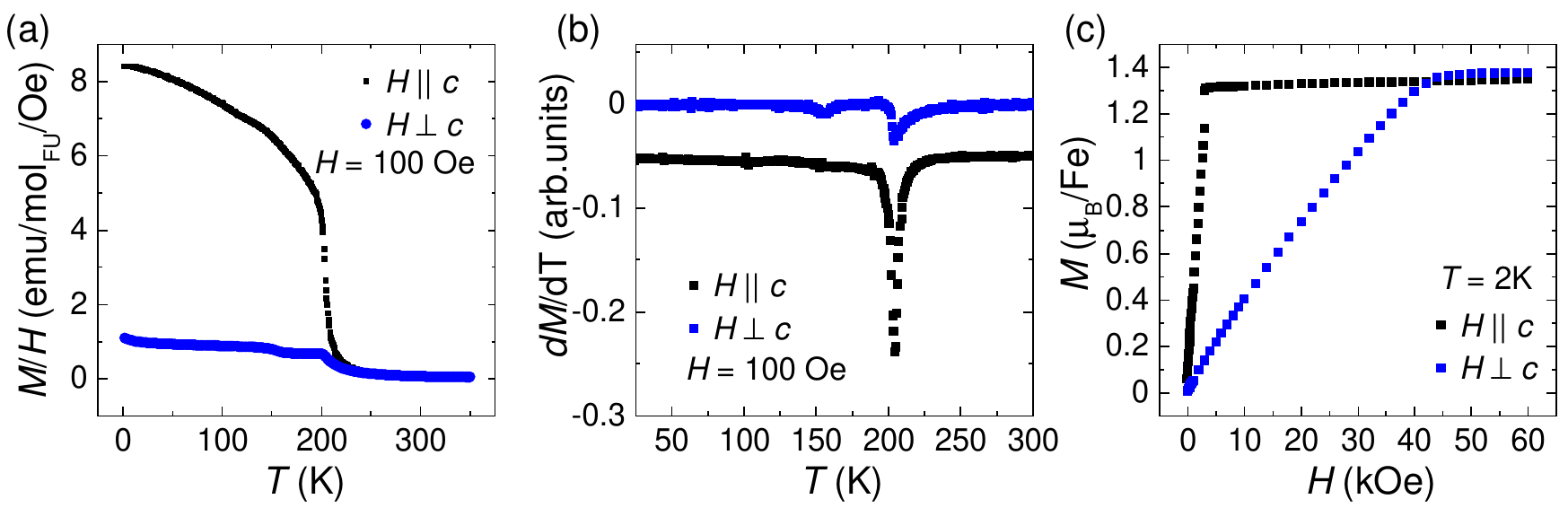}
\caption{The temperature and field dependence of bulk magnetization with external magnetic field applied parallel and perpendicular to the $c$-axis.  \label{fig:mag}}
\end{figure*}

\begin{figure*}[h]
\centering\includegraphics[width=0.8\textwidth]{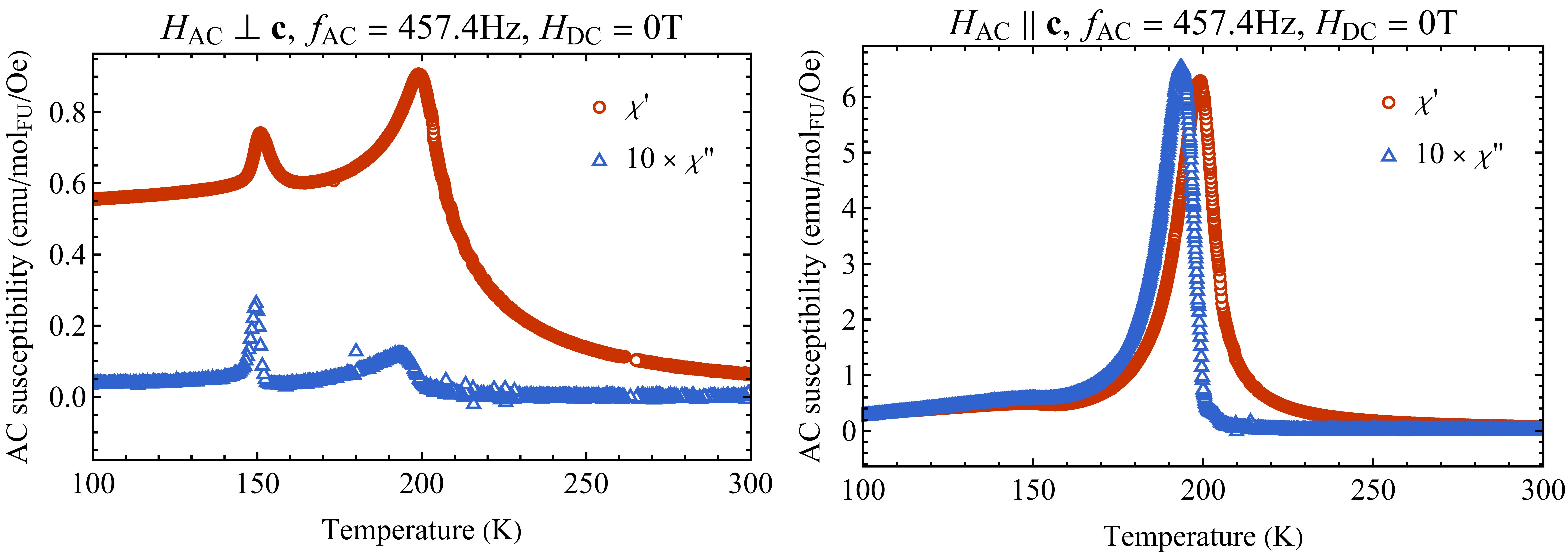}
\caption{The temperature dependence of AC susceptibility with zero external DC field. The data were collected from a crystal grown under the same condition as the 1g large crystal.
\label{fig:ac}}
\end{figure*}

\newpage
\section{Experimental details}

Elastic single-crystal neutron-scattering measurements were performed on the Elastic Diffuse Scattering Spectrometer (CORELLI) at the Spallation Neutron Source (SNS), Oak Ridge National Laboratory (ORNL), USA \citep{CORELLI}. The crystal was cooled down using a closed-cycle refrigerator (CCR). A full survey of reciprocal space was performed by rotating the sample over a range of $360^\circ$ with $2^\circ$ step at $T=5, 100$ and $250$\,K. The data was collected using a white beam and the elastic signals were reconstructed from the cross-correlation method \citep{CORELLI}. Every Q-position receives contributions from neutrons with a range of incident energies. The cross-correlation method provides an energy discrimination of $\Delta E/E_\text{i}$ = 3-5\%. For magnetic signals at low Q, dominant contributions come from neutrons with incident energies of $E_\text{i}$ = 12-50 meV. The corresponding energy resolution ranges from 0.4 to 2.5meV.\\

Inelastic neutron-scattering experiments were carried out on the Fine-Resolution Fermi Chopper Spectrometer (SEQUOIA) at SNS, ORNL \citep{SEQ}. Measurements at three temperatures $T=5, 155$ and $250$\,K were performed using an incident neutron energy $E_\text{i}=40$\,meV in the high resolution mode with an elastic resolution of FWHM $\approx 1.1$\,meV. The sample was rotated over a range of $360^\circ$ with $2^\circ$ step in each measurement. Additional data were collected at the base temperature $T=5$\,K with $E_\text{i}=25$\,meV in the high resolution mode (elastic FWHM $\approx 0.65$\,meV) and $E_\text{i}=150$ and $250$\,meV in the high flux mode (elastic FWHM $\approx 10$ and $17$\,meV respectively).\\

To track detailed temperature dependence of various features, complementary data were collected on the Triple-Axis Spectrometer (HB-3) at the High Flux Isotope Reactor (HFIR), ORNL. Measurements were performed with a fixed final energy $E_\text{f}=14.7$\,meV and a horizontal collimation 48'-40'-sample-40'-120', yielding elastic FWHM$\,\approx\,$0.8meV. Pyrolytic graphite (PG 002) monochromator and analyzer were used in the experiment. PG filters were placed between the sample and analyzer to reduce the contamination from higher-order scattering processes. \\

To further clarify whether the observed elastic peak at $Q_1=(0.3,0.3,0)$ is truly elastic or a result of integration of inelastic signals, additional scans were performed using Cold Neutron Triple-Axis Spectrometer (CTAX) at HFIR with $E_\text{f}=4$\,meV and a horizontal collimation open-80'-sample-open-open, yielding an elastic FWHM$\,\approx\,$0.2meV. Data is shown in the right panel of Fig.~S5. \\

Inelastic neutron-scattering experiments in magnetic fields were performed on the Wide Angular-Range Chopper Spectrometer (ARCS) at SNS, ORNL \citep{ARCS}. Magnetic fields were applied along the $c$-axis of the sample using Slim SAM 5T cryomagnet. Measurements were performed at $E_\text{i}=40$\,meV in the high resolution mode, yielding an elastic FWHM $\approx 1.68$\,meV.  Data were collected at $T=2.2$\,K with $\mu_0H$\,=\,$0$\,T and $4$\,T, covering a range of $180^\circ$ with $1^\circ$ step.\\

Raw neutron event data from time-of-flight experiments were converted into the histogram format using Mantid \citep{mantid}. Symmetrization according to the point group 6/mmm was applied to the histogram data to improve the statistics. Representative plots of the raw data are shown in Fig.\,\ref{fig:raw_DIFF} and \ref{fig:raw_INS}, where all the key observations presented in the main text can be unequivocally identified. The inelastic data were imported into Horace \citep{horace} and analyzed using a localized exchange interaction model $H $\,=\,$ \sum_{i<j}J_{ij}{\bf S}_i\cdot{\bf S}_i+D\sum_i(S^z_i)^2$ and SpinW \citep{spinw}. See Sec.~S5 for details of linear spin-wave modelling.

\begin{figure*}[h]
\centering\includegraphics[width=0.2\textwidth]{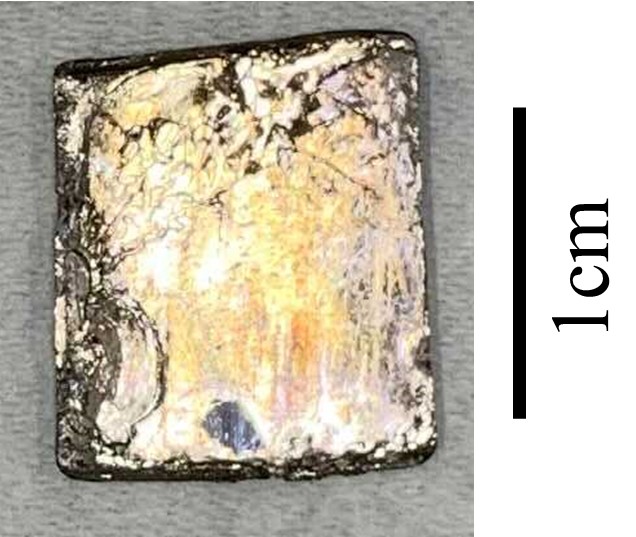}
\caption{1g large single crystal used in neutron scattering experiments. \label{fig:crystal}}
\end{figure*}

\newpage

\section{Additional neutron-scattering data}

\begin{figure*}[h]
\centering\includegraphics[width=0.95\textwidth]{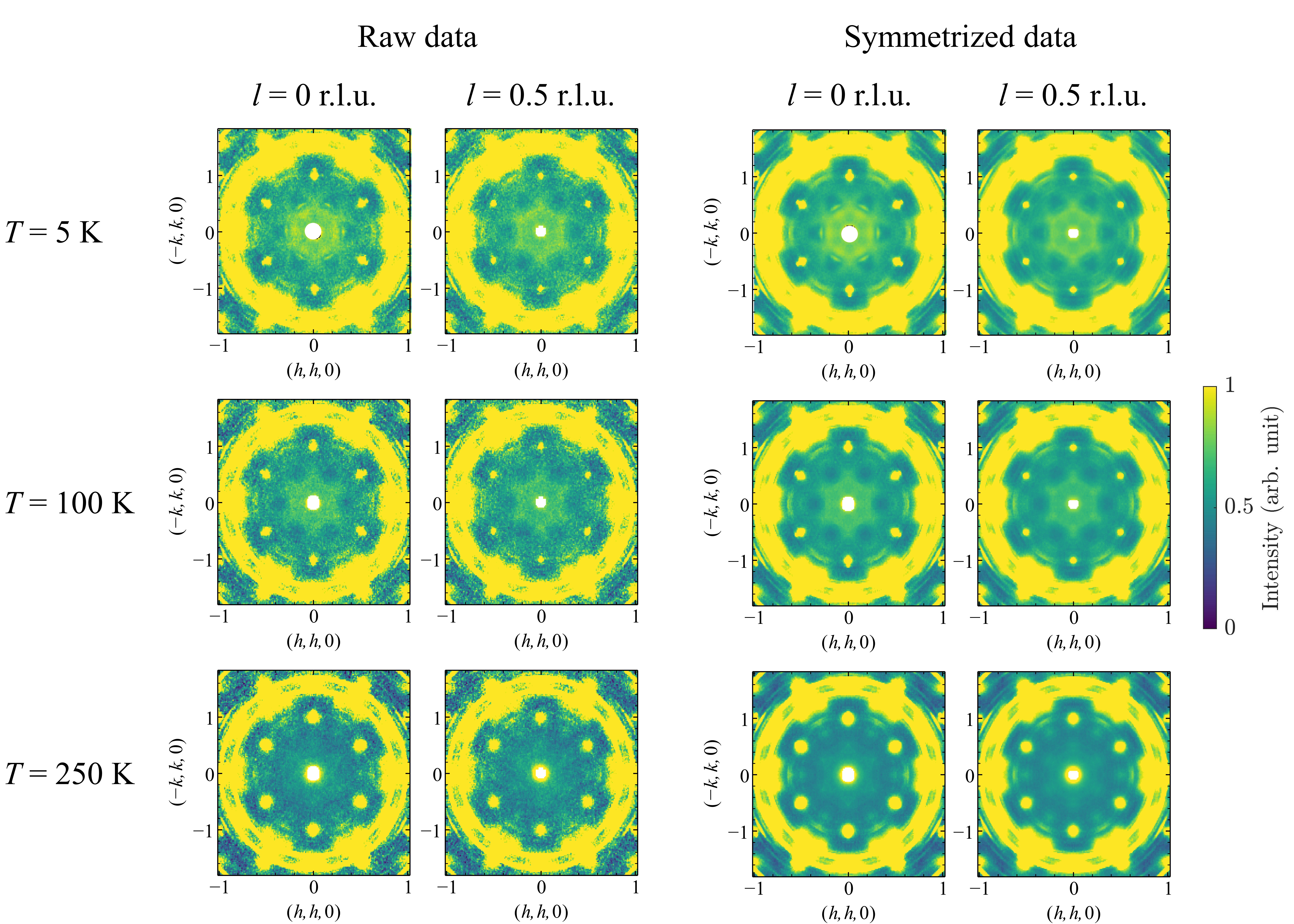}
\caption{Comparison between the raw and the symmetrized elastic neutron-scattering data measured on CORELLI. Integration of $\Delta l $\,=\,$ 0.1$ r.l.u. was performed for all cuts. Both features at ${\bf Q}_1$\,=\,$(0.3,0.3,0)$ and ${\bf Q}_2$\,=\,$(0.24,0.24,0)$ follow six-fold pattern in the raw data, therefore are intrinsic to the sample. \label{fig:raw_DIFF}}
\end{figure*}

\begin{figure*}[h]
\centering\includegraphics[width=0.35\textwidth]{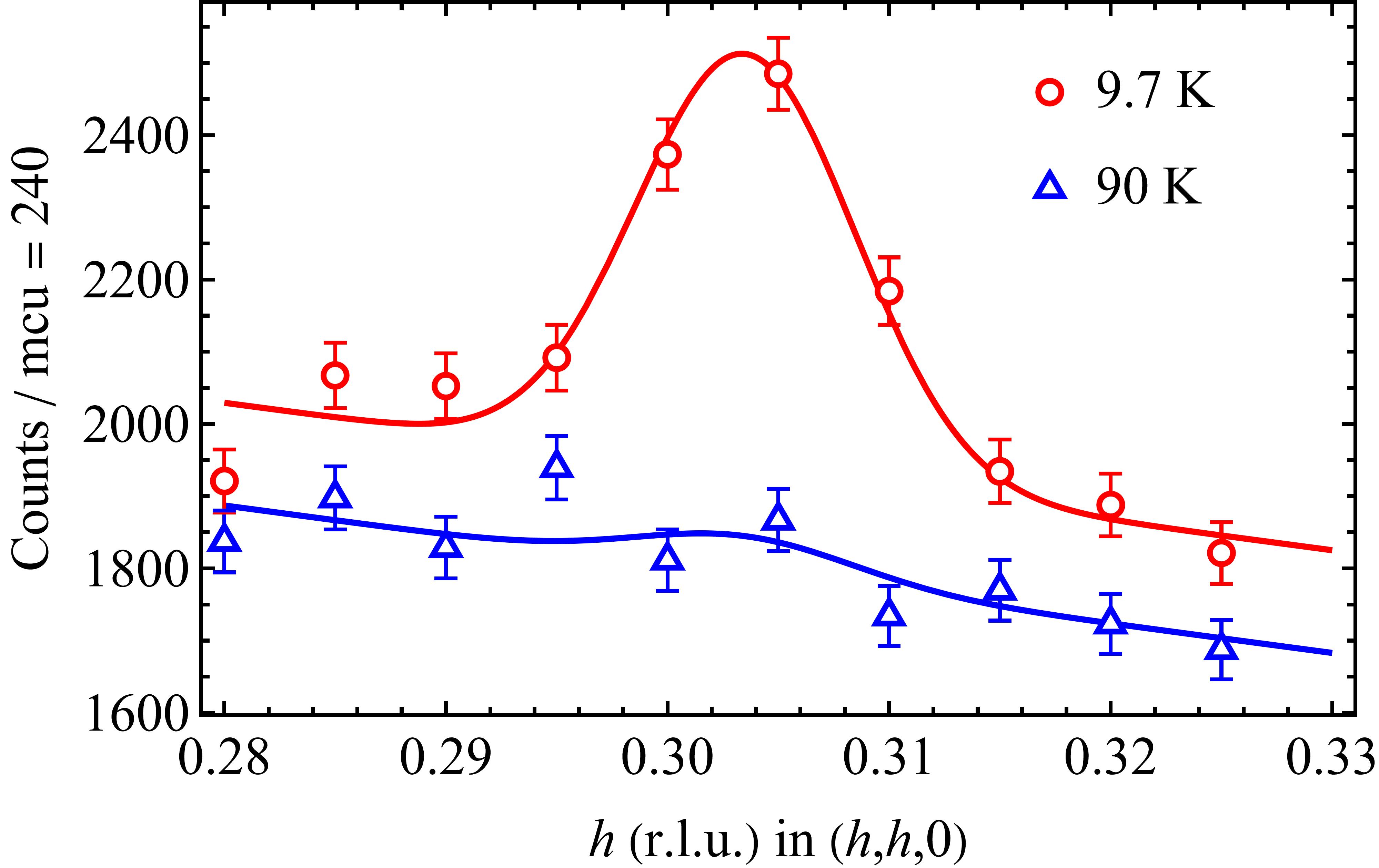}\hspace{1cm}\includegraphics[width=0.35\textwidth]{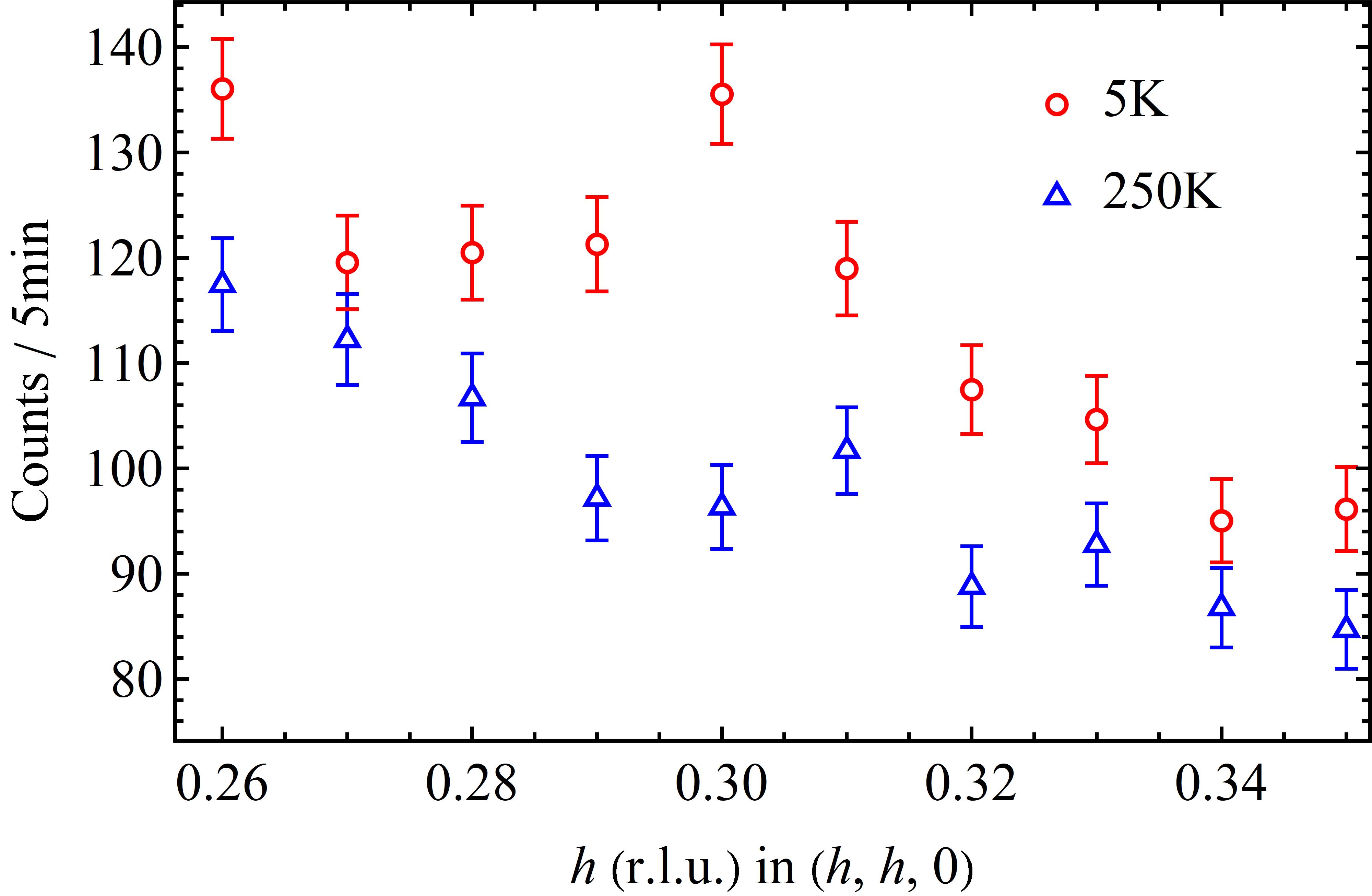}
\caption{Left: Longitudinal scans of the AFM Bragg at ${\bf Q}_1$\,=\,$(0.3,0.3,0)$ measured on HB-3 with FWHM$\,\approx\,$0.8meV. The lines are fits of a Gaussian peak with a linear background. The fitted heights of the Gaussian profiles are plotted in Fig.\,2(f). Right: The same scan performed on CTAX with a better elastic resolution (FWHM$\,\approx\,$0.2meV), which further confirms the presence of static AFM correlations. \label{fig:fit_OP}}
\end{figure*}

\newpage

\begin{figure*}[h]
\centering\includegraphics[width=0.95\textwidth]{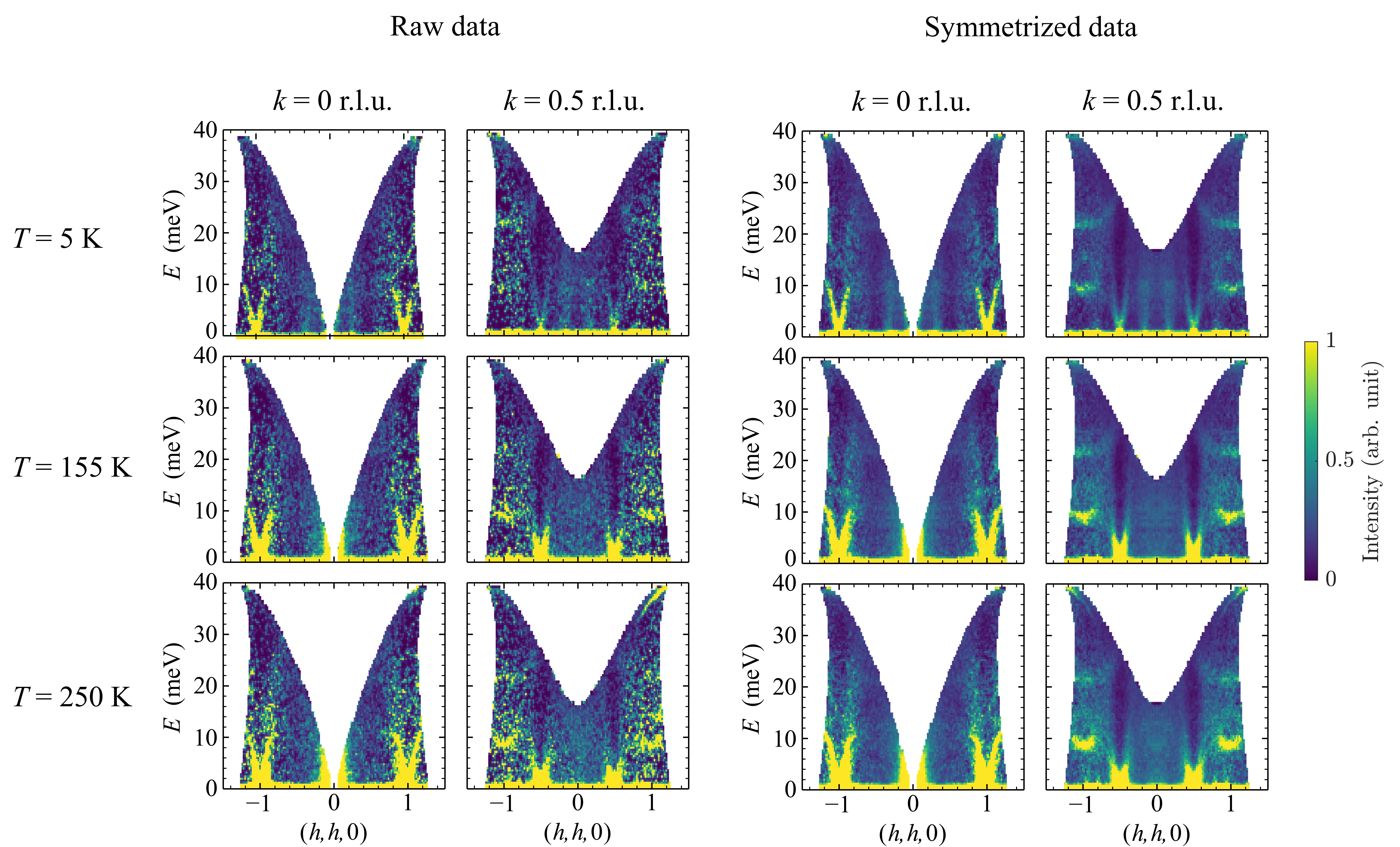}
\caption{Comparison between the raw and the symmetrized inelastic neutrons-scattering data measured on SEQUOIA. Integration of $\Delta l $\,=\,$ 1$ r.l.u. and $\Delta k $\,=\,$ 0.1$ r.l.u. was performed for all cuts. Different from Fig.\,2 where the Bose factor was divided out, raw scattering intensities are shown here. The AFM excitations emerge as broad incoherent signals around them diminish at low temperatures, therefore to track its temperature dependence, we focus on the M-point which is the place between the rod-like AFM excitations in Fig.\,\ref{fig:INS_OP}.
\label{fig:raw_INS}}
\end{figure*}

\begin{figure*}[h]
\centering\includegraphics[width=0.4\textwidth]{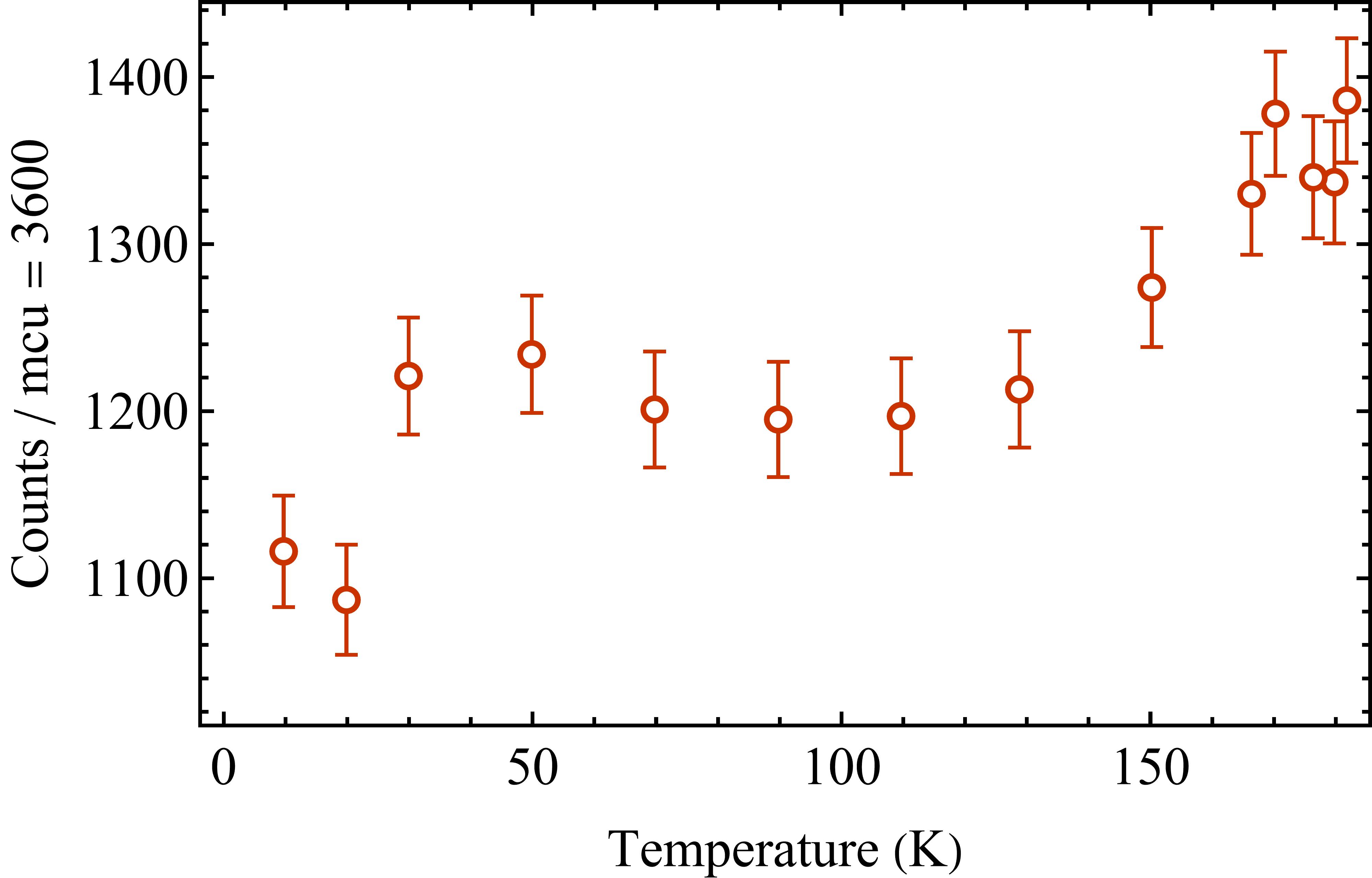}
\caption{Temperature dependence of inelastic neutron counts measured on HB-3 at the energy transfer $E$\,=\,$3$\,meV and ${\bf Q} $\,=\,$ (-0.5,0.5,0)$, the M-point. The neutron counts show a clear drop as temperature is decreased and start to flatten out below $\sim 130$\,K. This is likely to indicate the onset of the AFM excitations.  \label{fig:INS_OP}}\end{figure*}

\begin{figure*}[h]
\centering\includegraphics[width=1\textwidth]{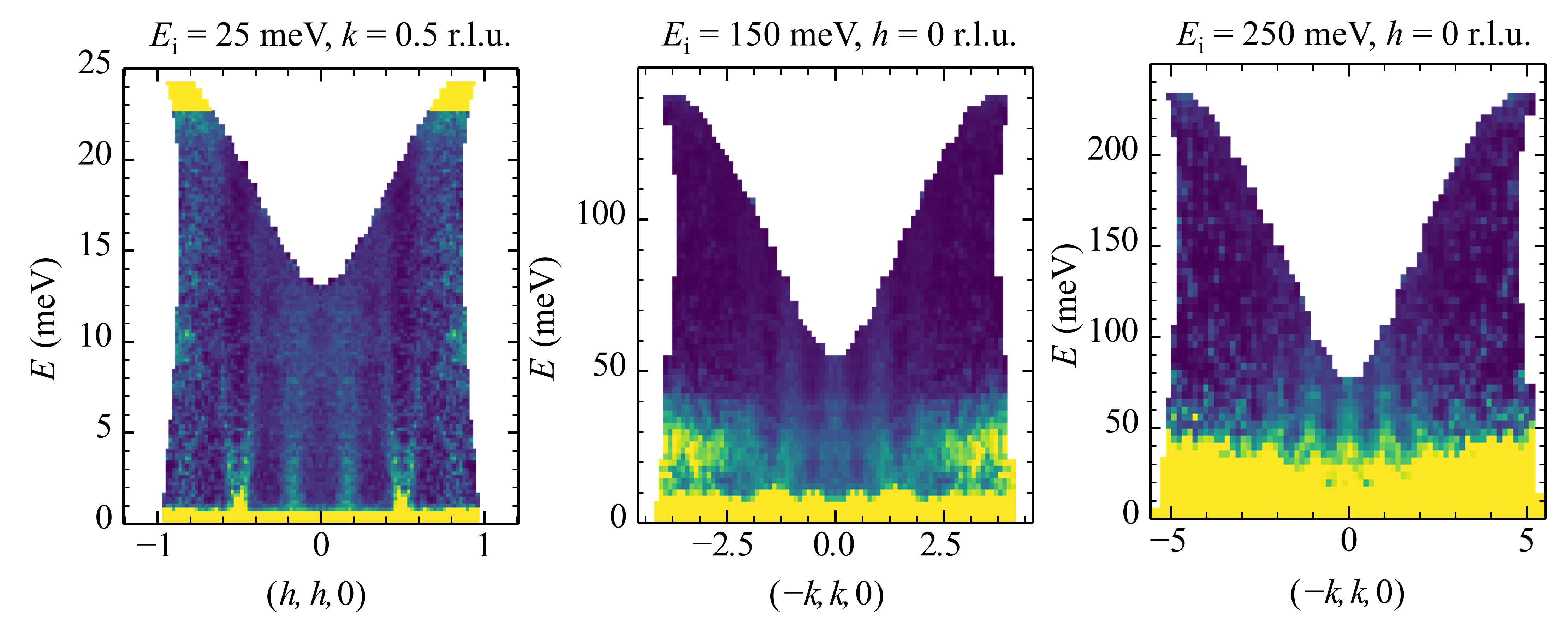}
\caption{Complementary inelastic neutron-scattering data collected on SEQUOIA at various incident energies. Integration of $\Delta l $\,=\,$ 1$ r.l.u. and $\Delta k (\Delta h) $\,=\,$ 0.1$ r.l.u. was performed for all cuts. \label{fig:other_Ei}}
\end{figure*}

\begin{figure*}[h]
\centering\includegraphics[width=0.8\textwidth]{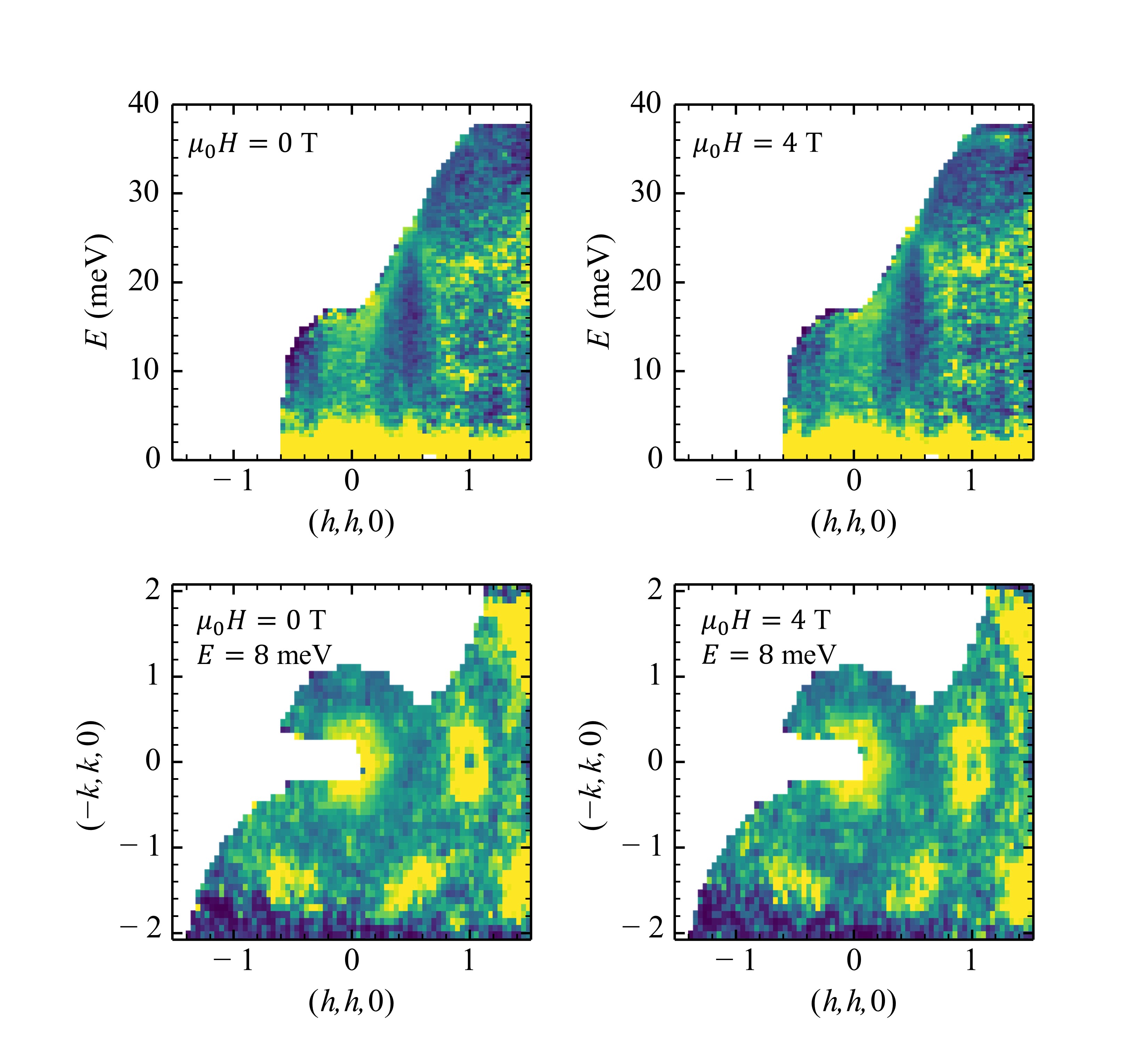}
\caption{Inelastic neutron scattering data measured on ARCS with external magnetic field applied along the $c$-axis. The momentum-energy cuts are integrated over $k $\,=\,$ 0.5\pm 0.05$\,r.l.u. and $l $\,=\,$ 0.0\pm 0.5$\,r.l.u.. The constant energy cuts are integrated over $E$\,=\,$8\pm1$\,meV. No clear difference can be seen between data in zero field and polarized phase at $\mu_0H$\,=\,$4$\,T. \label{fig:field_INS}}
\end{figure*}

\clearpage
\section{DFT+DMFT calculational setting and additional data}

The charge self-consistent DFT+DMFT scheme is used to describe FGT at the doping levels $\rm x=0.1,0.25$, based on the virtual-crystal approximation (VCA) applied to the Fe$^\text{II}$ site. 
For the DFT part, a mixed-basis pseudopotential 
framework~\citep{elsaesser90,lechermann02} in the generalized-gradient approximation (GGA) is put into practise. A $9\times 9\times 2$ k-point mesh is utilized and the plane-wave cutoff energy is set to $E_{\rm cut}=13$\,Ry. Local basis orbitals are introduced for
Fe$(3p,3d)$, Ge$(4s,4p)$ and Te$(5s,5p)$. The role of spin-orbit effects is neglected in the crystal calculations.

\begin{figure}[b]
\centering
(a)\includegraphics[width=0.6\columnwidth]{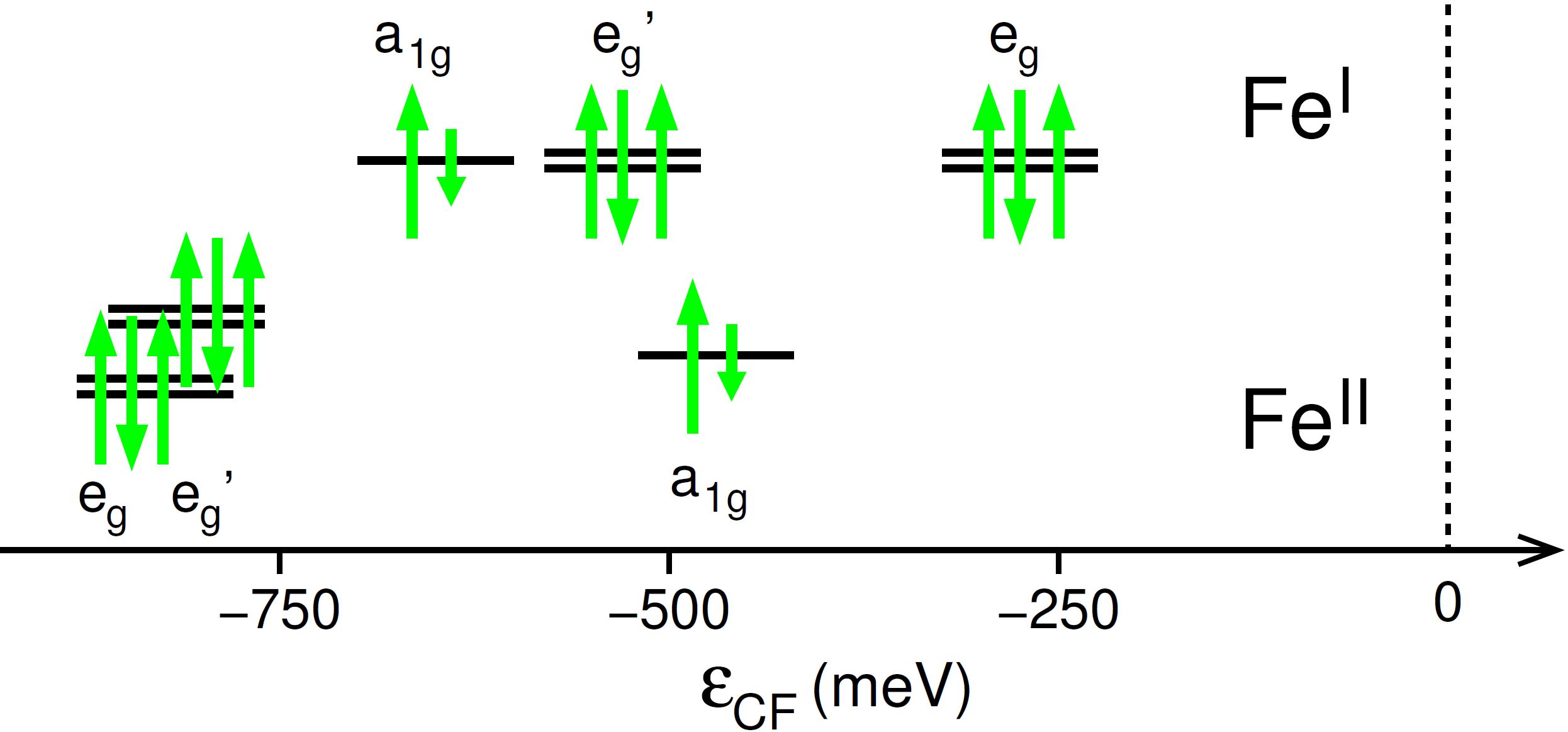}\\[0.2cm]
(b)\includegraphics[width=0.6\columnwidth]{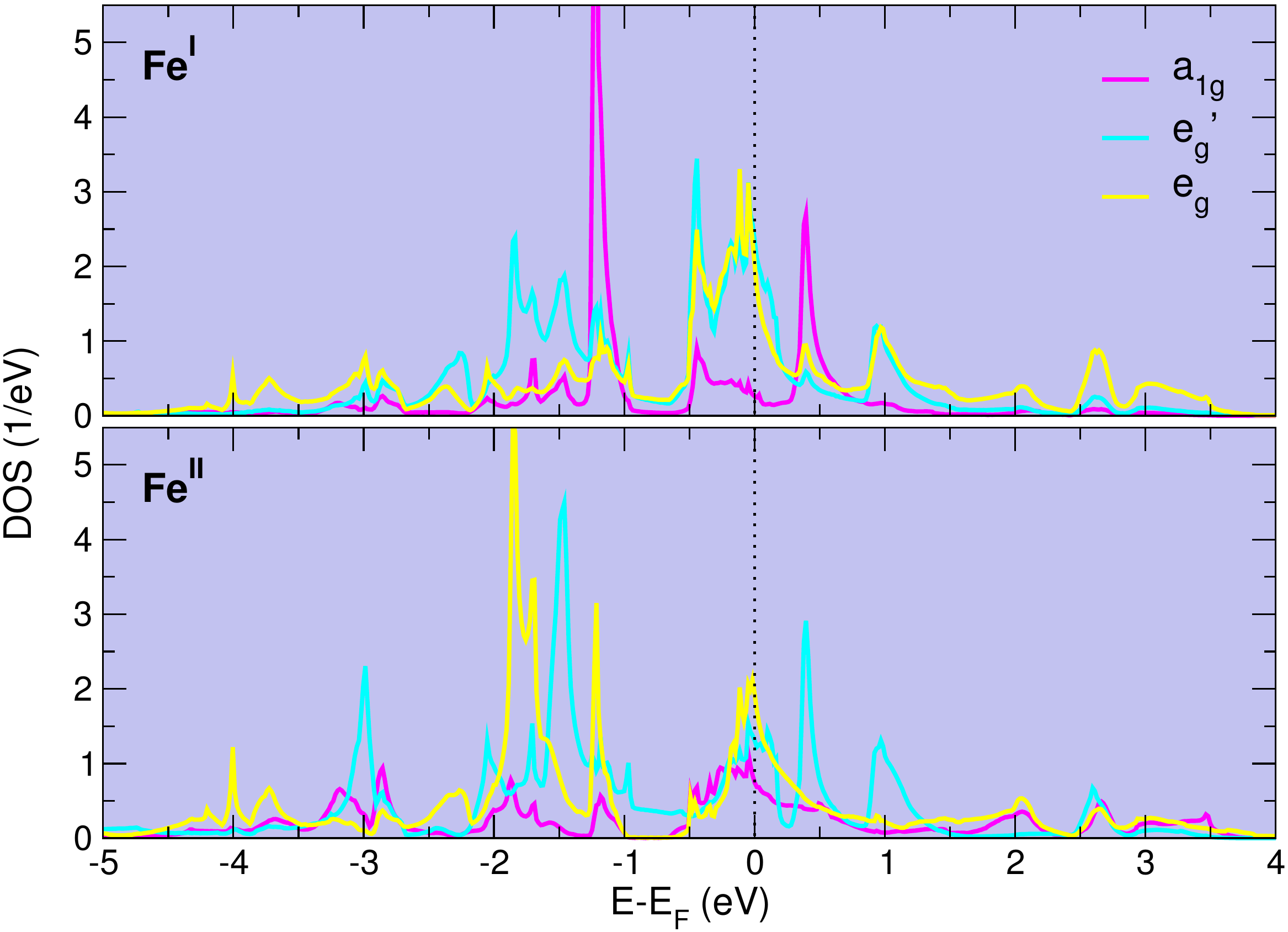}
\caption{\label{fig:OSMT_dft}  DFT data on the Fe$(3d)$ states in {\FGT}. (a) Crystal-field splitting of the two inequivalent Fe sites. The green arrows hint towards the principal spin-resolved occupation in the high-temperature phase.
(b) Site- and orbital-resolved Fe$(3d)$ density of states (DOS) for $\delta=0.1$. Top:  Fe$^\text{I}$, bottom: Fe$^\text{II}$.}
\end{figure}

The DMFT correlated subspace on each Fe site is governed by a Slater-Kanamori Hamiltonian, including density-density, spin-flip and pair-hopping terms. It is applied to the Fe$(3d)$ projected-local orbitals~\citep{amadon08}. The projection is performed on the Kohn-Sham bands above the 24 bands of dominant Fe$(3p)$, Ge$(4s)$ and Te$(5s)$ character. The projection window spans 50 bands, including the dominant KS states of dominant Fe$(3d)$, Ge$(4p)$ and Te$(5p)$ nature.
A Hubbard $U=5$\,eV and a Hund exchange $J_{\rm H}=0.7$\,eV are chosen to parametrize the local five-orbital Hamiltonian Slater-Kanamori Hamiltonian, respectively. The fully-localized-limit 
double-counting scheme~\cite{anisimov93} is applied. Continuous-time quantum Monte Carlo in hybridzation expansion~\cite{werner06} as implemented in the TRIQS code~\cite{parcollet15,seth16} is used to solve the multisite DMFT problem. Up to $1.5\cdot 10^{9}$ Monte-Carlo sweeps are performed to reach convergence. A Matsubara mesh of 1025(2049) frequencies is used to account for the higher(lower)-temperature regime. In detail, this means that 1025 frequencies are used for $T\ge$ 100\,K and 2049 frequencies are utilized for $T<$ 100\,K. For the analytical continuation from Matsubara space onto the real-frequency axis, the Maximum-entropy method~\cite{jarrell96} is used for the ${\bf k}$-integrated spectra (by continuation of the Bloch Green's function) and the Pad{\'e} method~\cite{vidberg77} is employed for the ${\bf k}$-resolved spectra (by continuation of the local self-energies).\\

On both symmetry-inequivalent Fe sites, the respective Fe$(3d)$ states in FGT split into three classes: a $d_{z^2}$-like $a_{1g}$ orbital, two degenerate $e_g'$ and two degenerate $e_g$ orbitals. The lobes of the $a_{1g}$ orbital point along the $c$-axis and are 'free' for the Fe$^\text{I}$ sites, but point towards the Te sites for the Fe$^\text{II}$ sites. The $e_{g}'$ orbitals formally complete with $a_{1g}$ the original $t_{2g}$ orbitals sector, and mainly point inbetween the neighboring sites. In contrast, the $e_g$ orbitals point more dominantly towards the neighboring sites. Both $e_{g}'$ and $e_{g}$ carry mixed in-plane and out-of-plane character. The DFT crystal-field (CF) levels, along with the sketched fillings, are pictorially presented in Fig. \ref{fig:OSMT_dft}(a). While the $a_{1g}$ orbitals are filled with close to 1.5 electrons, both other (twofold-degenerate) orbital sectors each carry about three electrons, respectively. The DFT density of states (DOS) for the $d$ states of Fe$^\text{I,II}$ are displayed in Fig. \ref{fig:OSMT_dft}(b) and reveal a strong bonding-antibonding splitting for Fe$^\text{I}$-$a_{1g}$. In general due to the large degree of covalency in the system, the Fe$(3d)$ weight is spread over a rather large energy range, not uncommon for Fe-based metals.\\

\begin{figure}[t]
\centering
\includegraphics[width=0.7\columnwidth]{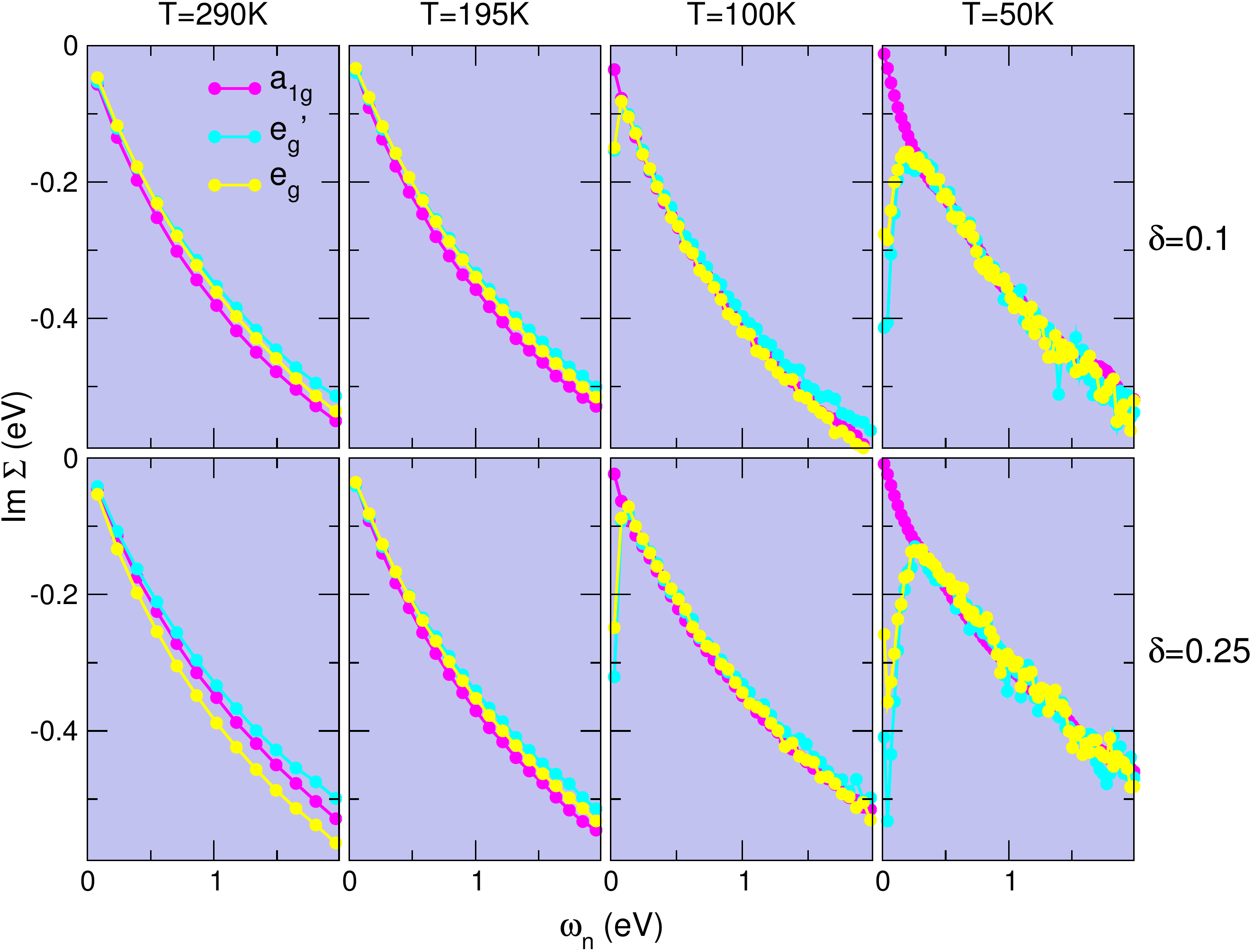}
\caption{\label{fig:OSMT_self} Imaginary part 
${\rm Im}\,\Sigma(i\omega_n)$ of the spin-averaged Fe$^\text{I}(3d)$ self-energy from DFT+DMFT at small Matsubara frequencies $\omega_n$. From ambient temperature down to 50\,K
(left to right); top row: $\delta=0.1$, bottom row: $\delta=0.25$.}
\end{figure}

Figure \ref{fig:OSMT_self} shows the temperature evolution of the imaginary part of the electron self-energy $\Sigma(i\omega_n)$ for Fe$^\text{I}(3d)$ in the range of small Matsubara frequencies. The OSMT regime at doping $\delta=0.1$ can be identified by the abrupt self-energy downturn in the $\{e_g',e_g\}$ orbital sector at low frequency for the data at $T=100$\,K, while the $a_{1g}$ self-energy remains Fermi-liquid-like with (near-)linear behavior towards zero frequency. A saturated bending of this downturn for $T=50$\,K marks the observed Kondo regime. Note that for $\delta=0.25$, the overall self-energy signature 
is very similar, however the downturn is somewhat smoother
and the $\{e_g',e_g\}$ response at 50\,K already renormalized metallic-like. The OSMT is nearly smeared out for the latter doping and the characteristics are closer to those of a strongly-renormalized metallic regime. For $\delta=0.1$, the $a_{1g}$ effective mass
$m^*/m_{\rm DFT}=1/Z=1-\partial\,{\rm Im}\,\Sigma/\partial\omega_n|_{\omega_n\rightarrow 0}$ results in moderate values of 1.6 at $T=290$\,K and of 2.0 at $T=50$\,K.\\

\begin{figure}[t]
\centering
\includegraphics[width=0.9\columnwidth]{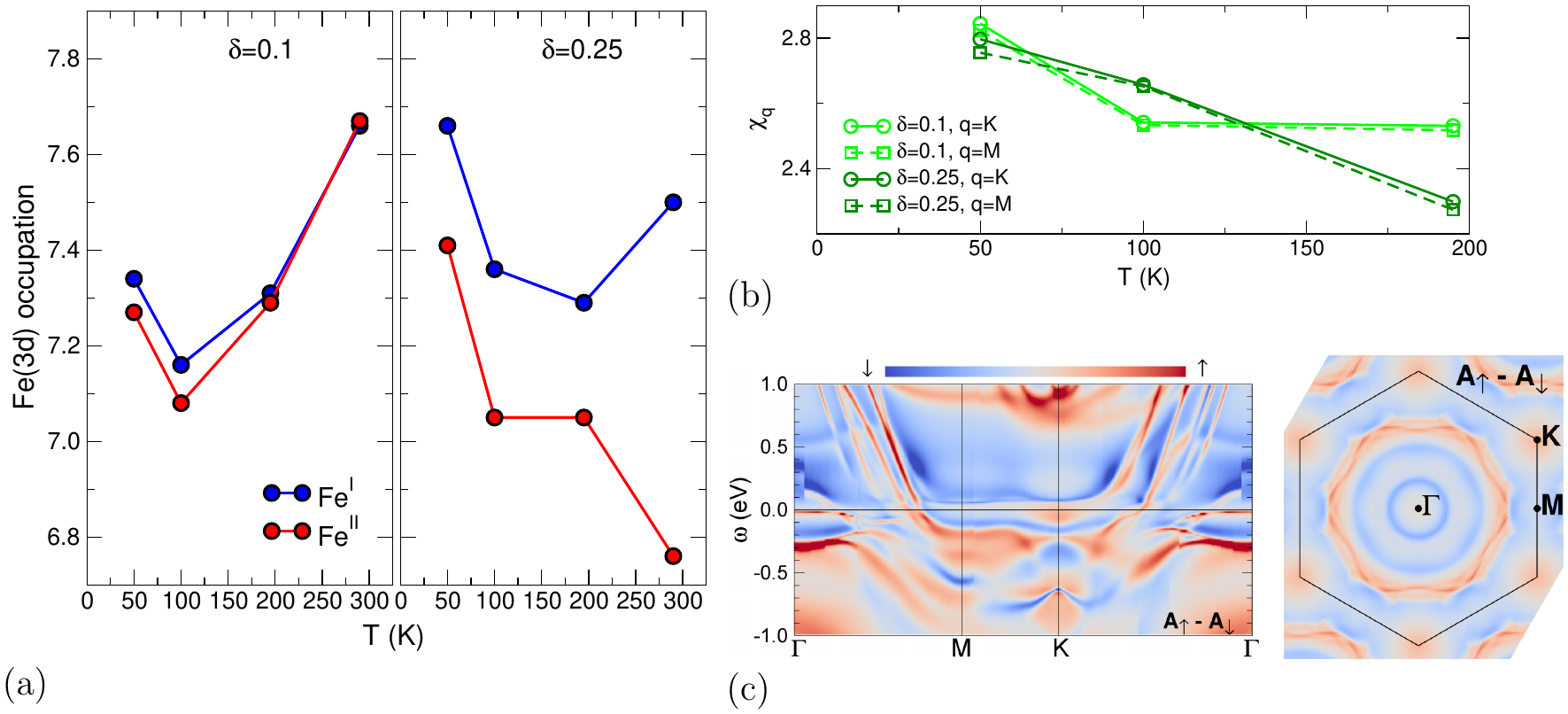}
\caption{\label{fig:dmftplus} Additional DFT+DMFT data on the physics of the Fe$(3d)$ states in {\FGT}. (a) Fe$(3d)$ occupation for Fe$^\text{I,II}$ with temperature for $\delta$=0.1 (left) and $\delta$=0.25 (right). 
(b) $T$-dependent Lindhard spin susceptibility $\chi_s^{(0)}({\bf q},\omega=0)$ for $\delta$=0.1,0.25 at K and M points, respectively. (c) {\bf k}- and spin-resolved spectral-function for $\delta=0.1$ at $T$=50\,K. $A_\uparrow({\bf k},\omega)-A_\downarrow({\bf k},\omega)$ along high-symmetry lines (left) and $k_z=0$ Fermi surface (right).}
\end{figure}

The correlated local Fe$(3d)$ charge, as obtained from the projected-local orbitals in DFT+DMFT, varies quite strongly with temperature as depicted in Fig.~\ref{fig:dmftplus}a. For $\delta$=0.1, the Fe$(3d)$ occupation decreases with $T$ and attains a minimum in the orbital-selective phase with about $\sim 7.1$ electrons. Thereby, the filling on the Fe$^\text{II}$ sites is always somewhat lower than on the Fe$^\text{I}$ sites. In the case of $\delta$=0.25, the charge differentiation between both symmetry-inequivalent Fe sites becomes substantial. The static part of the ${\bf q}$-dependent spin susceptibility $\chi_s^{(0)}$ is computed from the converged DFT+DMFT Green's function $G$ in the weak-coupling formulation 
$\chi_s^{(0)}({\bf q},0)\sim \sum_{{\bf k}n} G({\bf k},\omega_n)\,G({\bf k}+{\bf q},\omega_n)$ without including vertex corrections. The correlated nature of the {\FGT} FM phase is approached with spin-polarized DFT+DMFT calculations, whereby onlentry the DMFT part (i.e. the Fe$(3d)$ self-energy) carries the explicit spin polarization. This is the conventional scheme for magnetically-ordered phases, since an additional spin polarization in the DFT part usually leads to too strong exchange effects and an overestimation of Curie/N{\'e}el temperatures of interacting systems. However, the {\FGT} compound is not a `standard' correlated material (such as e.g. various transition-metal oxides), but metallic and with magnetism supposedly based also on subtle itinerant exchange. Therefore, it is not that surprising that the revealed 
site-resolved ferromagnetic Fe ordered moment of $m($Fe$^\text {I})=0.9$\,$\mu_{\rm B}$ and $m($Fe$^\text{ II})=0.6$\,$\mu_{\rm B}$ obtained at $T=100$\,K for $\delta=0.1$ is somewhat below the experimental estimates within the given theoretical approach. Figure~\ref{fig:dmftplus}b displays the $T$-dependent susceptibility data from Fig. 3b of the main text for the specific points $q$=K,M points at the zone boundary. It is seen that while $\chi_s^{(0)}$ grows at K for lower temperatures, it tends to saturate at M in this regime. As a proof of principles, the amplitude is also generally somewhat larger at the K point, in line with the experimental findings.
The spin-resolved spectral function and Fermi surface at $T$=50\,K is shown in Fig.~\ref{fig:dmftplus}c, marking the obvious spin dominance of certain Fermi-surface sheets. Note that the spectral intensity around the K point is weak and incoherent but not fully gapped.

\clearpage
\section{Spin-wave modelling for out-of-plane dispersions}
\begin{figure}[h!]
\centering\includegraphics[width=0.57\columnwidth]{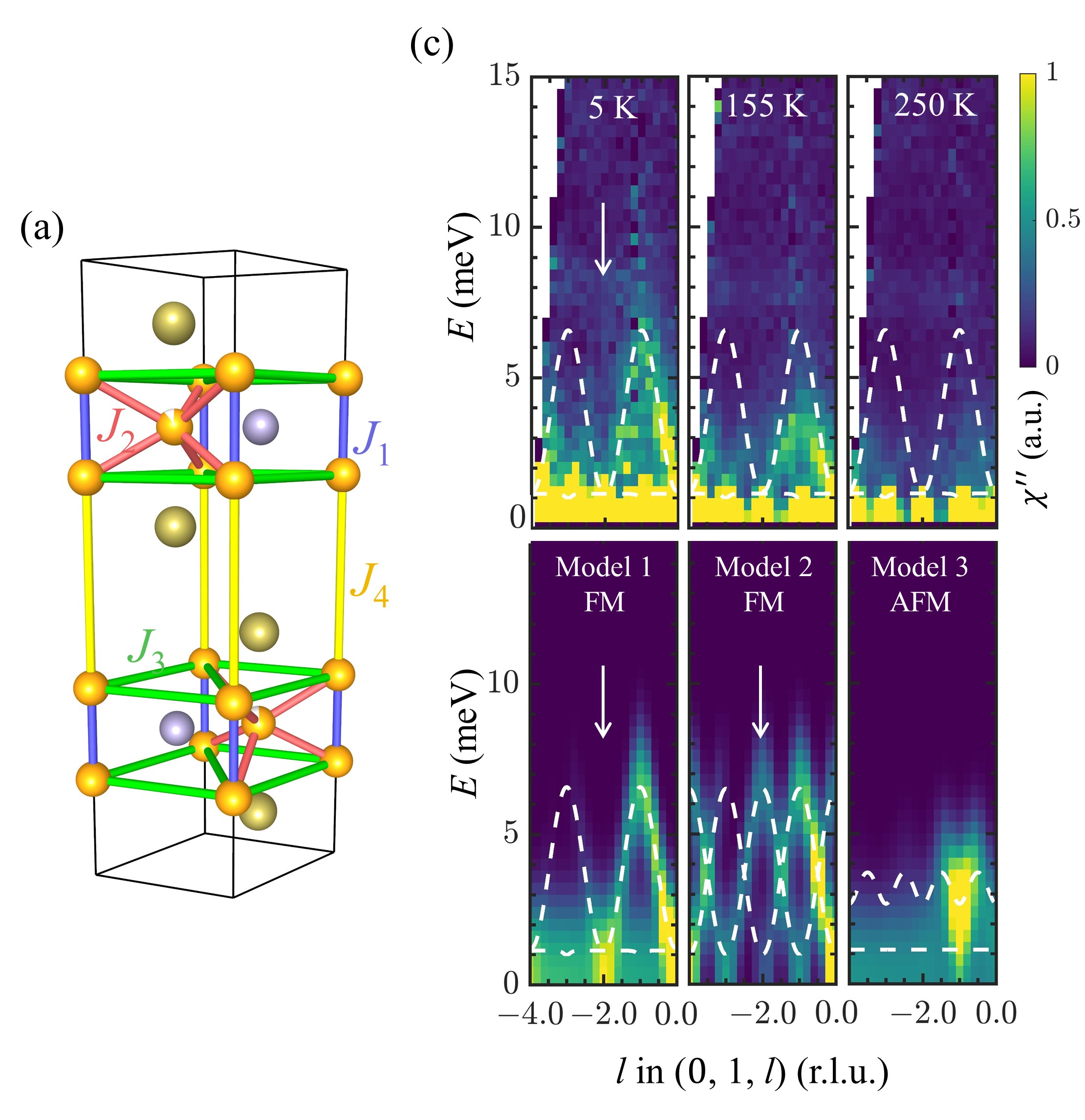}
\caption{\label{fig:sw} (a) Exchange pathways of FGT. (b) Top: temperature dependence of FM excitations in the out-of-plane directions at $k=0.5\pm0.05$ r.l.u. and $h=0.5\pm0.05$ r.l.u.. Bottom: calculations of three spin-wave models. The white arrow indicates the absence of intensities, which rules out Model 2 with a large $J_2$. See text for model parameters. The white dashed lines are dispersion bands with non-zero intensities calculated from Model 1. A broad energy-resolution (FWHM$\approx3$\,meV) is introduced to account for the damping effect.}
\end{figure}

Dispersive FM spin-waves are observed in the out-of-plane direction at $T$\,=\,$5$\,K, which is strongly temperature-renormalized at $155$\,K and completely disappear in the paramagnetic regime at $250$\,K [Fig.\,\ref{fig:sw}(c)]. This behavior
resembles the typical temperature dependence of magnetic excitations in insulating materials, reflecting the loss of coherence in the magnetic ground state due to increasing thermal fluctuations. In this case, it 
is associated with the destruction of long-range inter-layer correlations, revealing that the effective dimensionality is impacted by thermal fluctuations with evident 3D character at low temperatures. Nevertheless, the bandwidth of the out-of-plane dispersion is $\sim$7meV, more than a order of magnitude smaller than that of the in-plane excitations ($\sim$100\,meV), indicating dominant intra-layer interactions.  

This data allows us to improve on previous spin-wave models and make an estimation of the inter-layer coupling [Fig.\,\ref{fig:sw}(a)]. The $J_2$ bond connecting Fe$^\text{I}$ and Fe$^\text{II}$ was identified to be the dominant interaction in earlier neutron-scattering studies \citep{calder2019magnetic, trainer2021relating} and DFT calculations \citep{deng2018gate,jang2020origin}. Such a model, however, produces notable inelastic intensity at (0 1 $-$2), marked by the white arrow in Fig.\,\ref{fig:sw}(c) Model 2 ($J_1$\,=\,$J_2$\,=\,$-11$\,meV, $J_3$\,=\,$0$\,meV, $J_4$\,=\,$-1.7$\,meV), which is inconsistent with the data. To correct for it while still having a good description for in-plane dispersions, we keep $J_2$ to a minimum and introduce the $J_3$ bond that couples Fe$^\text{I}$ sites directly. The calculation is shown in the panel of Model 1 ($J_1$\,=\,$-11$\,meV, $J_2$\,=\,$-0.01$\,meV $J_3$\,=\,$-4.5$\,meV and $J_4$\,=\,$-1.1$\,meV), which agrees well with the data. This result suggests that the two Fe sites have very different characters and the Fe$^\text{II}$ sites may be magnetically more isolated than previously understood. Model 3 differs from Model 1 by changing $J_4$ to antiferromagnetic, yielding a poor description of the data, therefore further confirms the ferromagnetic inter-layer ordering. 
The easy-axis single-ion anisotropy is estimated from the saturation field ($H_\text{sat}\sim$\,4.4\,T) in the bulk magnetization data with field perpendicular to the $c$-axis. Using $D (S^z)^2 $\,=\,$ -g\mu_\text{B}SH_\text{sat}$ with $S^z$\,=\,$S$\,=\,$5/2$, $g$\,=\,$2$, $\mu_\text{B}$\,=\,$0.05788$\,meV/T and $H $\,=\,$ 4.4$\,T, we obtain $D$\,=\,$-0.2$\,meV. This is consistent with Ref.\,\citep{trainer2021relating} where $D$ is larger due to shorter spin length used. Spin length $S$\,=\,$5/2$ is assumed for both sites in our calculations. 

\clearpage
\section{Phonon  calculation}

In addition to the FM and AFM signals discussed in the main text, we also observe numerous low-energy modes that could be attributed to collective lattice vibrations, since their intensities become stronger with increasing temperature. To confirm the origin of such contributions to the inelastic neutron-scattering intensities, we obtain a DFT-calculated phonon spectra  and make a {\it qualitative} comparison with the experimental data in this section. Several factors need to be taken into account for a quantitative comparison (in future works), such as crystal defects, anharmonicity, electron/spin-phonon interaction, instrumental resolution and etc.

Spin-polarized DFT calculations of Fe$_3$GeTe$_2$ (P63/mmc, FM order) were performed using the Vienna Ab initio Simulation Package (VASP) \citep{kresse1996efficient}. The calculation used Projector Augmented Wave (PAW) method \citep{blochl1994projector,kresse1999ultrasoft} to describe the effects of core electrons, and Local Density Approximation (LDA) \citep{perdew1981self} for the exchange-correlation functional. Energy cutoff was $700$\,eV for the plane-wave basis of the valence electrons. The lattice parameters and atomic coordinates measured at $1.5$\,K \citep{zhuang2016strong} were used as the initial structure, and the structure was then relaxed to minimize the potential energy. The electronic structure was calculated on a $\Gamma$-centered mesh ($12\times12\times3$ for the unit-cell and $3\times3\times2$ for the supercell). The total energy tolerance for electronic energy minimization was $10^{-8}$ \,eV, and for structure optimization it was $10^{-7}$\,eV. The maximum interatomic force after relaxation was below $0.001$\, eV/\AA. The interatomic force constants were calculated by Density Functional Perturbation Theory (DFPT) on a $3\times3\times1$ supercell, and the vibrational eigen-frequencies and modes were then calculated using Phonopy \citep{togo2015first}. The OCLIMAX software \citep{cheng2019simulation} was used to convert the DFT-calculated phonon results to the simulated inelastic neutron-scattering spectra.

\begin{figure}[h!]
\centering\includegraphics[width=1\columnwidth]{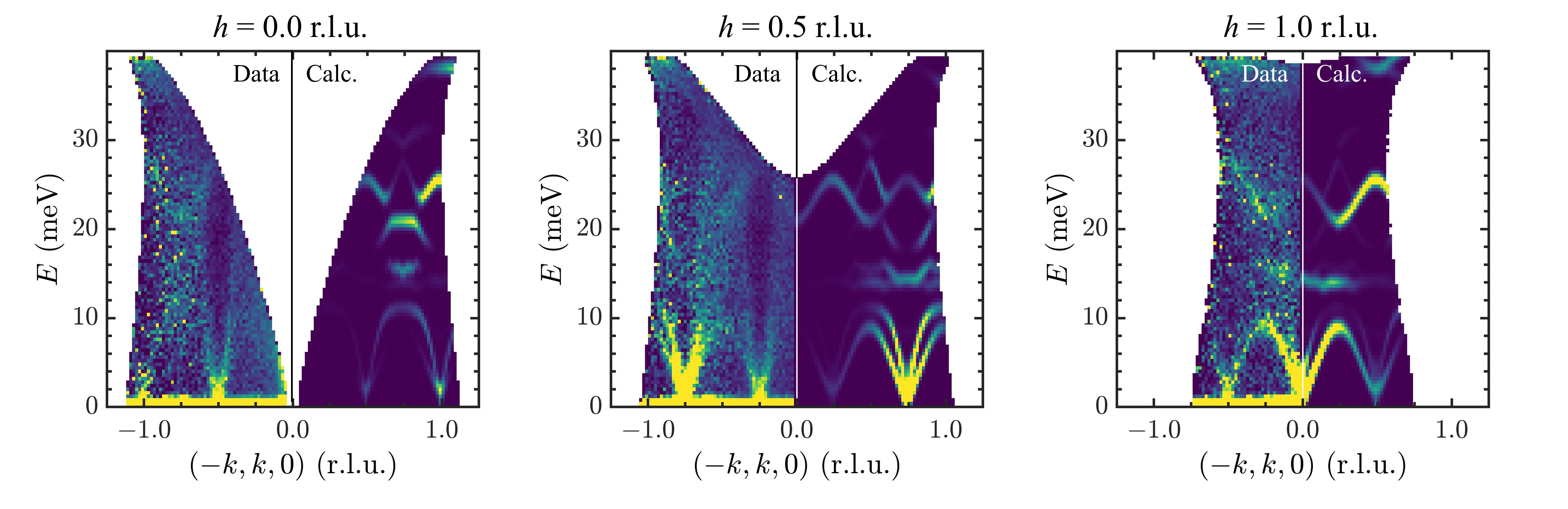}
\caption{\label{fig:phonon} Comparison between inelastic neutron data measured on SEQOUIA at $T=5$~K and phonon calculation. All cuts are integrated over $\Delta h = 0.1$\,r.l.u. and $l = 0.
0\pm 0.5$\,r.l.u..}
\end{figure} 


\end{document}